\documentclass{article}
\usepackage{arxiv}
\usepackage[utf8]{inputenc}
\usepackage[T1]{fontenc}
\usepackage{amsmath}
\usepackage{etoolbox}
\AtBeginEnvironment{equation}{}
\AtBeginEnvironment{align}{}
\usepackage{txfonts}
\usepackage{multicol}
\usepackage{enumitem}
\usepackage{float}
\usepackage{graphicx}
\usepackage{tikz}
\usetikzlibrary{shapes, shapes.geometric, arrows, chains, positioning}
\usepackage{epstopdf}
\usepackage{booktabs}
\usepackage{cite}

\usepackage[hidelinks]{hyperref}
\title{GPU Acceleration of Collinear and Noncollinear DFT Using a Numerical Atomic Orbital-Based DFT Code}
\renewcommand{\undertitle}{}
\renewcommand{\shorttitle}{GPU Acceleration of Collinear and Noncollinear DFT}
\date{\footnotesize Published in J. Phys. Soc. Jpn. \textbf{94}, 124003 (2025), \url{https://doi.org/10.7566/JPSJ.94.124003}}

\author{%
  Hiroyuki Kawai\,$^{1}$\thanks{Corresponding author: \texttt{h.river1234@gmail.com}}\quad
  Takuya Sekikawa\,$^{2}$\quad
  Taisuke Ozaki\,$^{3}$\quad
  Yoshiaki \=Ono\,$^{1}$ \\[2pt]
  \normalfont\normalsize $^{1}$Department of Physics, Niigata University, Niigata 950-2181, Japan \\
  \normalfont\normalsize $^{2}$Nuclear Science and Engineering Center, Japan Atomic Energy Agency, Tokai, Ibaraki 319-1195, Japan \\
  \normalfont\normalsize $^{3}$Institute for Solid State Physics, The University of Tokyo, Kashiwa, Chiba 277-8581, Japan%
}

\begin{document}
\maketitle

\begin{abstract}
We implement GPU acceleration of collinear and noncollinear density functional theory (DFT) calculations in the numerical atomic orbitals (NAOs) code OpenMX by offloading matrix multiplications and eigenvalue solves (plus selected auxiliary steps) to cuBLAS/cuSOLVER and OpenACC. 
Benchmarks on the Pegasus supercomputer (per node: a 48-core Intel Xeon Platinum 8468 CPU and one NVIDIA H100 GPU) compare GPU-accelerated and CPU-only runs under identical settings. 
For a 512-atom collinear case on two nodes (two GPUs total), the GPU-accelerated calculation achieves a 2.02 times speedup over a CPU-only run on two nodes (96 CPU cores total); for a 384-atom noncollinear case on two nodes (two GPUs total), the speedup is 2.60 times over two CPU-only nodes (96 cores). 
These results demonstrate practical GPU-accelerated DFT in an NAO-based code for both collinear and noncollinear calculations.
\end{abstract}


\section{Introduction}
A graphics processing unit (GPU) is a specialized processor that provides high parallel processing capability and wide memory bandwidth~\cite{gpu, gpu2, gpuac}. In high-performance computing (HPC), systems combining CPUs and GPUs are common because they offer high computational performance and excellent energy efficiency~\cite{gpu3, gpu4, gpuac}. Although GPUs can accelerate many numerical simulations, their characteristics are not universally advantageous, and not all simulations benefit from GPU acceleration~\cite{gpuac2, gpu5}. Nevertheless, Kohn--Sham density-functional theory (DFT) calculations have achieved significant success using GPU technology~\cite{gpuacgaussian1, gpuacgaussian2, gpuacgamess_uk, gpuacnwchem, gpuacbigdft1, gpuacbigdft2, gpuacfhiaims, gpuacsiesta, gpuacabacus, gpuacopenmx}.

DFT~\cite{dft1, dft2} is widely used across various disciplines to predict the properties of materials and molecules~\cite{dftuse1, dftuse2, dftuse3, dftuse4}. While the plane-wave pseudopotential (PW) method~\cite{pw1, pw2, pw3, pw4, pw5} is the most common implementation, approaches employing spatially localized atomic orbitals (AOs) are also utilized. These include Gaussian-type orbitals~\cite{gto1, gto2, gto3, gto4, gto5, gto6, gto7}, Slater-type orbitals~\cite{sto1, sto2}, and numerical atomic orbitals (NAOs)~\cite{nao1, nao2, nao3, nao4, nao5, nao6, nao7, nao8, nao9, nao10, nao11, nao12, nao13}. The localization of AOs allows for many local operations on a real-space grid, such as evaluating integrals for Hamiltonian and overlap matrix elements, updating electron density, and computing forces and stress tensors~\cite{gpuacfhiaims, dftao}. For small systems (e.g., tens of atoms), these real-space grid operations dominate the computational cost~\cite{nao12, gpuacfhiaims}. However, for larger systems (hundreds to thousands of atoms), the dominant cost shifts to solving the Kohn--Sham eigenvalue problem~\cite{dftao2, gpuacfhiaims}. The time required for this step scales as $O(N^3)$ with the system size $N$~\cite{dftao, dftao2}. Recently, efforts have focused on accelerating the Kohn--Sham eigenvalue problem using GPUs~\cite{gpuacsiesta, gpuacabacus, gpuacopenmx}.

GPU acceleration has been reported in several AO-based DFT codes, including Gaussian~\cite{gpuacgaussian1, gpuacgaussian2}, GAMESS-UK~\cite{gpuacgamess_uk}, NWChem~\cite{gpuacnwchem}, BigDFT~\cite{gpuacbigdft1, gpuacbigdft2}, FHI-aims~\cite{gpuacfhiaims}, SIESTA~\cite{gpuacsiesta}, ABACUS~\cite{gpuacabacus}, and OpenMX~\cite{gpuacopenmx}. For medium- to large-scale systems, the strategy for porting these codes to GPUs focuses on reducing the cost of the most intensive parts of the generalized Kohn--Sham eigenvalue problem—namely, matrix multiplication and eigenvalue solving—by reimplementing them on GPUs. To the best of our knowledge, while GPU acceleration for collinear DFT in NAO-based codes has been reported, it has not been reported for noncollinear DFT, which allows spin orientation to vary three-dimensionally~\cite{ncdft1, ncdft2, ncdft3, ncdft4}. Existing GPU implementations for OpenMX~\cite{nao9,nao10,nao11} are limited to collinear DFT~\cite{gpuacopenmx}. Therefore, in this work, we implement GPU acceleration in OpenMX for both collinear and noncollinear DFT.

\section{DFT Calculation with NAO}
For simplicity, we assume a non-spin-polarized system in this section; the extension to spin-polarized cases is straightforward, and noncollinear spin will be discussed later. In DFT, the total-energy functional of the ground state is given by
\begin{equation}
\begin{split}
E_{\mathrm{KS}}\left[  n\right]  =T_{\mathrm{s}}[n]+E_{\mathrm{ext}}[n]
      +E_{\mathrm{H}}[n]+E_{\mathrm{xc}}[n]+E_{\mathrm{nuc\text{-}nuc}},
\end{split}
\label{eq:total_ene}
\end{equation}
where $n$ is the electron density, $T_\mathrm{s}$ the kinetic energy of the non-interacting reference system, $E_\mathrm{ext}$ the external energy (generally from electron-nucleus attraction), $E_\mathrm{H}$ the Hartree energy, $E_\mathrm{xc}$ the exchange--correlation energy, and $E_\mathrm{nuc\text{-}nuc}$ the nucleus--nucleus repulsion energy.
The ground-state density $n_{0}(\mathbf{r})$ (and thus the ground-state total energy) is obtained by minimizing Eq.~\eqref{eq:total_ene} under the constraint of a fixed number of electrons $N$. This variational procedure yields the Kohn--Sham (KS) equation:
\begin{equation}
\hat{H}_{\mathrm{KS}}\psi_{i}
=\left[\hat{t}_{\mathrm{s}}+v_{\mathrm{ext}}+v_{\mathrm{H}}+v_{\mathrm{xc}}\right]\psi_{i}
=\varepsilon_{i}\psi_{i},
\label{eq:ks_equ}
\end{equation}
where $\hat{H}_{\mathrm{KS}}$ is the KS Hamiltonian, $\hat{t}_{\mathrm{s}}$ the kinetic-energy operator of the non-interacting system, $v_{\mathrm{ext}}$ the external potential, $v_{\mathrm{H}}$ the Hartree potential, $v_{\mathrm{xc}}$ the exchange--correlation potential, and $\varepsilon_{i}$ the eigenvalue of the $i$-th one-electron orbital.
The electron density is expressed as
\begin{equation}
n\!\left(\mathbf{r}\right)=
2\sum_{i}f_{i}\,|\psi_{i}(\mathbf{r})|^{2},
\end{equation}
where the summation runs over all orbitals, and $f_i$ denotes the Fermi--Dirac distribution function.
Because $\hat{H}_{\mathrm{KS}}$ depends on $n(\mathbf{r})$, Eq.~\eqref{eq:ks_equ} is a nonlinear equation. Finding the stationary point of Eq.~\eqref{eq:total_ene} is equivalent to solving Eq.~\eqref{eq:ks_equ} to identify a Hamiltonian for which the input and output electron densities coincide. This nonlinear minimization problem is solved using the self-consistent-field (SCF) method.

To solve Eq.~\eqref{eq:ks_equ} numerically, we expand each KS orbital in a finite basis set $\{\chi_{m}\}$:
\begin{equation}
\psi_{i}(\mathbf{r})=
\sum_{m=1}^{N_{b}}c_{mi}\,\chi_{m}(\mathbf{r}),
\label{eq:expansion}
\end{equation}
where a finite nonperiodic system is assumed, $N_{b}$ is the number of basis functions, and $c_{mi}$ the $m$-th component of the $i$-th eigenvector. Substituting Eq.~\eqref{eq:expansion} into Eq.~\eqref{eq:ks_equ} leads to the generalized eigenvalue problem
\begin{equation}
\mathbf{H}\mathbf{C}=\mathbf{S}\mathbf{C}\mathbf{\varepsilon},
\end{equation}
where $\mathbf{H}$ is the Hamiltonian matrix, $\mathbf{S}$ the overlap matrix, and $\mathbf{C}$ contains the eigenvectors. $\mathbf{S}$ arises because the NAOs do not form an orthonormal set. Each element $S_{lm}$ of $\mathbf{S}$ is defined as
\begin{equation}
S_{lm}=%
{\displaystyle\int}
d\mathbf{r}\,\chi_{l}^{\ast}\left(  \mathbf{r}\right)  \chi
_{m}\left(  \mathbf{r}\right).
\end{equation}
The electron density can then be calculated as
\begin{equation}
n\left(  \mathbf{r}\right)  =2%
{\displaystyle\sum\limits_{l,m=1}^{N_{b}}}
\chi_{l}^{\ast}\left(  \mathbf{r}\right)  n_{lm}%
\chi_{m}\left(  \mathbf{r}\right),
\end{equation}
where $n_{lm}$ denotes an element of the density matrix, defined as
\begin{equation}
n_{lm}=%
{\displaystyle\sum\limits_{i}}
f_{i}c_{li}^{\ast}c_{mi}.
\end{equation}

For periodic systems, the KS orbital with crystal momentum $\mathbf{k}$ is expanded in Bloch-summed basis functions $\{\tau_{m}^{(\mathbf{k})}\}$:
\begin{equation}
\Psi_{i}^{(\mathbf{k})}(\mathbf{r})=
\sum_{m=1}^{N_{b}}c_{mi}^{(\mathbf{k})}\,
\tau_{m}^{(\mathbf{k})}(\mathbf{r}),
\label{eq:psi}
\end{equation}
where $\mathbf{k}$ is the crystal-momentum quantum number and $N_b$ the number of basis functions in the unit cell.
Each Bloch-sum basis function $\tau_{m}^{\left(  \mathbf{k}\right)  }$ is written in terms of the numerical atomic orbital $\chi_m$ and the lattice vector $\mathbf{R}_n$ as
\begin{equation}
\tau_{m}^{\left(  \mathbf{k}\right)  }\left(  \mathbf{r}\right)
=\dfrac{1}{\sqrt{N}}%
{\displaystyle\sum\limits_{\mathbf{R}_{n}}}
e^{i\mathbf{k\cdot R}_{n}}\chi_{m}\left(  \mathbf{r-R}%
_{n}\right),
\label{eq:tau}
\end{equation}
where $N$ is the total number of unit cells under the Born--von K\'{a}rm\'{a}n periodic boundary condition.
Using Eqs.~\eqref{eq:psi} and \eqref{eq:tau}, Eq.~\eqref{eq:ks_equ} is rewritten as a generalized eigenvalue problem:
\begin{equation}
\mathbf{H}^{\left(  \mathbf{k}\right)  }\mathbf{C}^{\left(  \mathbf{k}\right)
}=\mathbf{S}^{\left(
\mathbf{k}\right)  }\mathbf{C}^{\left(  \mathbf{k}\right)}\mathbf{\varepsilon}^{\left(  \mathbf{k}\right)  }.
\label{eq:eigenval_problem}
\end{equation}
The element $S_{lm}^{\left( \mathbf{k}\right)}$ of the overlap matrix $\mathbf{S}^{(\mathbf{k})}$ is defined as
\begin{equation}
\begin{aligned}
S_{lm}^{\left(  \mathbf{k}\right)  }&=%
{\displaystyle\int}
d\mathbf{r}\tau_{l}^{\left(  \mathbf{k}\right)  \ast}\left(
\mathbf{r}\right)  \tau_{m}^{\left(  \mathbf{k}\right)  }\left(
\mathbf{r}\right) \\
&=%
{\displaystyle\sum\limits_{\mathbf{R}_{n}}}
e^{i\mathbf{k\cdot R}_{n}}%
{\displaystyle\int}
d\mathbf{r\chi}_{l}^{\ast}\left(  \mathbf{r}\right)  \chi
_{m}\left(  \mathbf{r+R}_{n}\right).
\end{aligned}
\end{equation}
We now briefly present the formulation for noncollinear DFT. The equation corresponding to Eq.~\eqref{eq:ks_equ} is
\begin{equation}
\begin{aligned}
\hat{H}_{\mathrm{NC\text{-}KS}}\left(
\begin{array}
[c]{c}%
\psi_{i}^{\alpha}\\
\psi_{i}^{\beta}%
\end{array}
\right)  =\varepsilon_{i}\left(
\begin{array}
[c]{c}%
\psi_{i}^{\alpha}\\
\psi_{i}^{\beta}%
\end{array}
\right)
\end{aligned}
\label{eq:ks_equ_nc}
\end{equation}
with
\begin{equation}
\begin{aligned}
\hat{H}_{\mathrm{NC\text{-}KS}}=\left(
\begin{array}
[c]{cc}%
\hat{t}_{\mathrm{s}}+v_{\mathrm{ext}}^{\alpha\alpha}+v_{\mathrm{H}%
}+v_{\mathrm{xc}}^{\alpha\alpha} & v_{\mathrm{ext}}^{\alpha\beta
}+v_{\mathrm{xc}}^{\alpha\beta}\\
v_{\mathrm{ext}}^{\beta\alpha}+v_{\mathrm{xc}}^{\beta\alpha} & \hat
{t}_{\mathrm{s}}+v_{\mathrm{ext}}^{\beta\beta}+v_{\mathrm{H}}+v_{\mathrm{xc}%
}^{\beta\beta}%
\end{array}
\right),
\end{aligned}
\end{equation}
where $\psi^{\alpha}$ and $\psi^{\beta}$ are the $\alpha$ and $\beta$ components of the spinor wave function.
The exchange--correlation potential is
\begin{equation}
v_{\mathrm{xc}}^{\sigma\sigma^{\prime}}=\dfrac{\delta E_{\mathrm{xc}}%
[n]}{\delta n^{\sigma\sigma^{\prime}}},
\end{equation}
where $\sigma$ denotes the $\alpha$- or $\beta$-component. The quantities $v^{\alpha\alpha}_{\mathrm{ext}}$, $v^{\alpha\beta}_{\mathrm{ext}}$, $v^{\beta\alpha}_{\mathrm{ext}}$, and $v^{\beta\beta}_{\mathrm{ext}}$ are the corresponding components of the external potential.
As in the collinear case, we expand \(\psi_{i}^{\sigma}\) in Eq.~\eqref{eq:ks_equ_nc} using a finite set of basis functions \(\chi(\mathbf{r})\):
\begin{equation}
\psi_{i}^{\sigma}\left(  \mathbf{r}\right)  =%
{\displaystyle\sum\limits_{m=1}^{N_{b}}}
c_{mi}^{\sigma}\chi_{m}\left(  \mathbf{r}\right).
\end{equation}
By defining the following matrix elements:
{%
\small
\begin{alignat}{2}
& H_{lm}^{\alpha\alpha} &&= \int d\mathbf{r}\,
  \chi_{l}^{\ast}(\mathbf{r})
  \left(\hat t_{\mathrm{s}}+v_{\mathrm{ext}}^{\alpha\alpha}+v_{\mathrm{H}}+v_{\mathrm{xc}}^{\alpha\alpha}\right)
  \chi_{m}(\mathbf{r}), \\
& H_{lm}^{\alpha\beta}  &&= \int d\mathbf{r}\,
  \chi_{l}^{\ast}(\mathbf{r})
  \left(v_{\mathrm{ext}}^{\alpha\beta}+v_{\mathrm{xc}}^{\alpha\beta}\right)
  \chi_{m}(\mathbf{r}), \\
& H_{lm}^{\beta\alpha}  &&= \int d\mathbf{r}\,
  \chi_{l}^{\ast}(\mathbf{r})
  \left(v_{\mathrm{ext}}^{\beta\alpha}+v_{\mathrm{xc}}^{\beta\alpha}\right)
  \chi_{m}(\mathbf{r}), \\
& H_{lm}^{\beta\beta}  &&= \int d\mathbf{r}\,
  \chi_{l}^{\ast}(\mathbf{r})
  \left(\hat t_{\mathrm{s}}+v_{\mathrm{ext}}^{\beta\beta}+v_{\mathrm{H}}+v_{\mathrm{xc}}^{\beta\beta}\right)
  \chi_{m}(\mathbf{r}) .
\end{alignat}
}%
Equation~\eqref{eq:ks_equ_nc} reduces to the generalized eigenvalue problem:
\begin{align}
\left(
\begin{array}
[c]{cc}%
\mathbf{H}^{\alpha\alpha} & \mathbf{H}^{\alpha\beta}\\
\mathbf{H}^{\beta\alpha} & \mathbf{H}^{\beta\beta}%
\end{array}
\right)  \left(
\begin{array}
[c]{c}%
\mathbf{C}^{\alpha}\\
\mathbf{C}^{\beta}%
\end{array}
\right)
& \nonumber \\
&\!\!\!\!\!\!\!\!\!\!\!\!\!\!\!\!\!\!\!\!\!\!\!\!\!\!\!\!\!\!\!\!\!\!\!\! = \left(
\begin{array}{cc}
\mathbf{S} & \mathbf{0} \\
\mathbf{0} & \mathbf{S}
\end{array}
\right)
\left(
\begin{array}{c}
\mathbf{C}^{\alpha} \\
\mathbf{C}^{\beta}
\end{array}
\right)\mathbf{\varepsilon}.
\end{align}
The noncollinear electron density \(n_{\sigma\sigma'}(\mathbf{r})\) is given by
\begin{align}
\left(
\begin{array}
[c]{cc}%
n^{\alpha\alpha}\left(  \mathbf{r}\right)   & n^{\alpha\beta}\left(
\mathbf{r}\right)  \\
n^{\beta\alpha}\left(  \mathbf{r}\right)   & n^{\beta\beta}\left(
\mathbf{r}\right)
\end{array}
\right) & \nonumber \\
&\!\!\!\!\!\!\!\!\!\!\!\!\!\!\!\!\!\!\!\!\!\!\!\!\!\!\!\!\!\!\!\!\!\!\!\!\!\!\!\!\!\!\!\!\! =
{\displaystyle\sum\limits_{l,m=1}^{N_{b}}}
\chi_{l}^{\ast}\left(  \mathbf{r}\right)  \left(
\begin{array}
[c]{cc}%
n_{lm}^{\alpha\alpha} & n_{lm}^{\alpha\beta}\\
n_{lm}^{\beta\alpha} & n_{lm}^{\beta\beta}%
\end{array}
\right)  \chi_{m}\left(  \mathbf{r}\right),
\end{align}
where \(n^{\sigma\sigma^{\prime}}_{lm}\) are the elements of the noncollinear density matrix, defined as
\begin{equation}
n_{lm}^{\sigma\sigma^{\prime}}=%
{\displaystyle\sum\limits_{i}}
f_{i}c_{li}^{\sigma\ast}c_{mi}^{\sigma^{\prime}%
}.
\end{equation}
The up-spin electron density \(n^{\uparrow}(\mathbf{r})\) and the down-spin electron density \(n^{\downarrow}(\mathbf{r})\) are computed as follows:
\begin{align}
\left(
\begin{array}
[c]{cc}%
n^{\uparrow}\left(  \mathbf{r}\right)   & 0\\
0 & n^{\downarrow}\left(  \mathbf{r}\right)
\end{array}
\right) & \nonumber \\
&\!\!\!\!\!\!\!\!\!\!\!\!\!\!\!\!\!\!\!\!\!\!\!\!\!\!\!\!\!\!\!\!\!\!\!\!\!\!\!\!\!\! =
{\displaystyle\sum\limits_{l,m=1}^{N_{b}}}
\chi_{l}^{\ast}\left(  \mathbf{r}\right)  \mathbf{U}\left(
\begin{array}
[c]{cc}%
n_{lm}^{\alpha\alpha} & n_{lm}^{\alpha\beta}\\
n_{lm}^{\beta\alpha} & n_{lm}^{\beta\beta}%
\end{array}
\right)  \mathbf{U}^{\dag}\chi_{m}\left(  \mathbf{r}\right),
\end{align}
where \(\mathbf{U}\) is the unitary matrix that diagonalizes the noncollinear density matrix, and \(\dagger\) denotes the conjugate transpose.

\section{Implementation of GPU Acceleration}
We implement GPU acceleration in OpenMX version 3.9.9 by modifying the routines invoked during diagonalization: \texttt{Band\_DFT\_Col.c} (collinear DFT, Band\_DFT\_Col routine) and \texttt{Band\_DFT\_NonCol.c} (noncollinear DFT, Band\_DFT\_NonCol routine).

\subsection{GPU acceleration of collinear DFT}
\begin{figure}[bt]
  \centering
  \tikzstyle{block} = [rectangle, draw, fill=blue!20,
      text width=17em, text centered, rounded corners, minimum height=4em]
  \tikzstyle{line} = [draw, -latex', thick]
\begin{tikzpicture}[auto]
  \node [block] (part1) {Part 1. For each $\mathbf{k}$-point, construct the overlap matrix $\mathbf{S}$ and the Hamiltonian matrix $\mathbf{H}$.};
  \node [block, below=1cm of part1] (part2) {Part 2. Solve the generalized eigenvalue problem $\mathbf{H}\mathbf{C} = \mathbf{S}\mathbf{C}\mathbf{\varepsilon}$ and obtain the eigenvalues $\varepsilon$ and eigenvectors $\mathbf{C}$.};
  \node [block, below=1cm of part2] (part3) {Part 3. Compute the chemical potential and the band energies.};
  \node [block, below=1cm of part3] (part4) {Part 4. Compute the charge-density and energy-density matrices.};
  \path[line] (part1) -- (part2);
  \path[line] (part2) -- (part3);
  \path[line] (part3) -- (part4);
\end{tikzpicture}
\caption{Computational flow of the Band\_DFT\_Col routine.}
\label{fig:flowchart_col}
\end{figure}
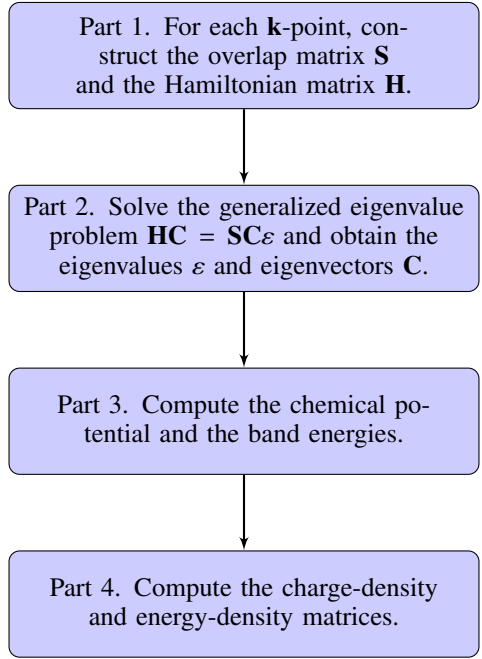

Figure~\ref{fig:flowchart_col} shows the flowchart for the Band\_DFT\_Col routine used for diagonalization in collinear DFT. OpenMX employs Löwdin orthogonalization to transform the generalized eigenvalue problem into a standard eigenvalue problem.

The computations in Parts 1 and 2 loop over assigned \(\mathbf{k}\)-points. Although Parts 1 through 4 include steps other than solving the generalized eigenvalue problem, OpenMX collectively refers to this sequence as “diagonalization”.
The loop over $\mathbf{k}$-points is parallelized across Message Passing Interface (MPI) processes.
If the number of MPI processes exceeds the number of $\mathbf{k}$-points to be calculated, the loop over $\mathbf{k}$-points is eliminated, and the computation for a single $\mathbf{k}$-point is distributed among multiple MPI processes.
In conventional (CPU) OpenMX, Part 1 is MPI-parallelized, generating $\mathbf{S}$ and $\mathbf{H}$ in a block-cyclic distribution across MPI ranks. However, in our GPU-accelerated implementation, Part 1 is not MPI-parallelized. This is because parallelizing Part 1 would require gathering the block-cyclic-distributed matrices to a single MPI rank for each $\mathbf{k}$-point before proceeding to Part 2. 
Due to the high cost of MPI communication involved in this aggregation process, we do not employ this approach in our work.
Furthermore, Part 1, which is not MPI-parallelized as will be discussed later, is computationally expensive (see Table~\ref{tab:breakdown_scf_gpu}) yet difficult to accelerate with GPUs, and thus has not been GPU-accelerated. A straightforward GPU implementation of the deeply nested loops involved in the numerical integration for constructing these matrices would require extensive atomic operations, resulting in poor performance.

\begin{figure}[bt]
  \centering
  \tikzstyle{block} = [rectangle, draw, fill=blue!20,
      text width=17em, text centered, rounded corners, minimum height=4em]
  \tikzstyle{line} = [draw, -latex', thick]
  \begin{tikzpicture}[auto]
      \node [block] (part21) {Part 2-1. Diagonalize $\mathbf{S}$ to obtain a unitary matrix $\mathbf{U}$ and a diagonal matrix $\mathbf{s}$ such that $\mathbf{U}^{\dagger}\mathbf{S}\mathbf{U}=\mathbf{s}$.};
      \node [block, below=1cm of part21] (part22) {Part 2-2. Construct $\mathbf{V}=\mathbf{U}\mathbf{s}^{-1/2}$.};
      \node [block, below=1cm of part22] (part23) {Part 2-3. Form $\mathbf{H}'=\mathbf{V}^{\dagger}\mathbf{H}\mathbf{V}$.};
      \node [block, below=1cm of part23] (part24) {Part 2-4. Diagonalize $\mathbf{H}'$ to obtain eigenvalues $\varepsilon$ and eigenvectors $\mathbf{C}'$.};
      \node [block, below=1cm of part24] (part25) {Part 2-5. Recover the eigenvectors of $\mathbf{H}$ as $\mathbf{C}=\mathbf{V}\mathbf{C}'$.};
      \path[line] (part21) -- (part22);
      \path[line] (part22) -- (part23);
      \path[line] (part23) -- (part24);
      \path[line] (part24) -- (part25);
  \end{tikzpicture}
  \caption{Detailed flow of Part 2 in the Band\_DFT\_Col routine.}
  \label{fig:flowchart_col_part2}
\end{figure}
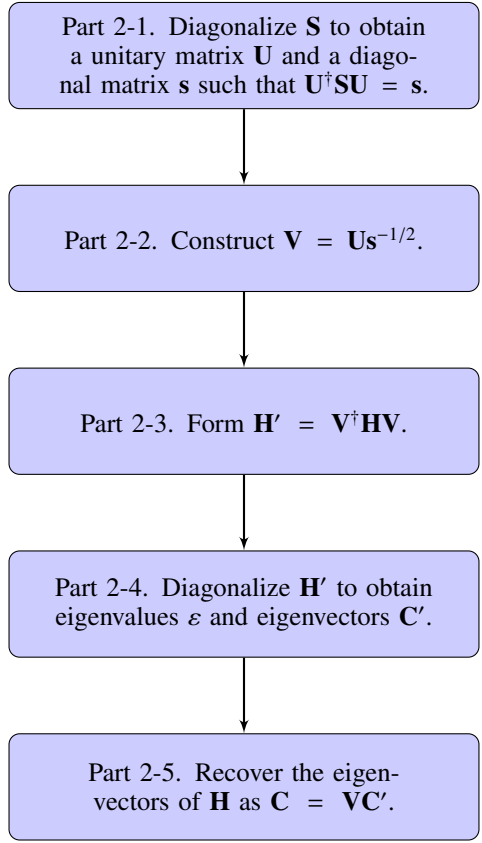

The procedure in Part 2 of the Band\_DFT\_Col routine is subdivided as shown in Fig.~\ref{fig:flowchart_col_part2}.

In Part 2-1, $\mathbf{U}$ is the unitary matrix that diagonalizes $\mathbf{S}$, and $\mathbf{s}$ is the diagonalized form of $\mathbf{S}$. The matrix dimension equals the total number of basis functions in the system.

In this study, we accelerate Part 2 on the GPU by replacing ScaLAPACK~\cite{scalapack} with cuBLAS~\cite{cuBLAS} and ELPA~\cite{elpa1, elpa2} with cuSOLVER~\cite{cuSOLVER}.
ScaLAPACK is parallelized with MPI, and the ELPA library is also highly MPI-parallelized~\cite{elpa1, elpa2}.
Unlike ScaLAPACK or ELPA, which rely on block-cyclic distribution across MPI processes, cuBLAS and cuSOLVER operate directly on the full matrix held by a single MPI process. Consequently, even when multiple MPI processes handle one $\mathbf{k}$-point, the GPU-accelerated portions use only one MPI process.

We therefore disable MPI parallelization from Part 1 through Part 2-5 and assign the full matrix to a single MPI process. We replace ELPA with cuSOLVER in Parts 2-1 and 2-4, and ScaLAPACK with cuBLAS in Parts 2-3 and 2-5. Part 2-2 is GPU-accelerated using OpenACC. Although this step does not utilize Tensor Cores, it accounts for a negligible portion of the runtime, as discussed later.
Procedures from Part 3 onward require block-cyclically distributed matrices. Thus, at the end of Part 2-5, we redistribute the full matrix held by one MPI process into a block-cyclic layout across all MPI processes. We do not implement GPU acceleration for Part 3 because its computational cost is negligible compared to the total diagonalization time (see Table~\ref{tab:breakdown_scf_gpu}). GPU acceleration of Part 4 is challenging despite its relatively higher cost (see Table~\ref{tab:breakdown_scf_gpu}); its large memory requirements exceed the available device memory, and the presence of deeply nested loops necessitates atomic operations in a straightforward implementation, degrading performance. Therefore, these stages remain MPI-parallelized using the original OpenMX source code. The replacement of ScaLAPACK/ELPA with cuBLAS/cuSOLVER necessitates these modifications to the Band\_DFT\_Col routine.

\subsection{GPU acceleration of noncollinear DFT}
\begin{figure}[bt]
  \centering
  \tikzstyle{block} = [rectangle, draw, fill=blue!20,
      text width=17em, text centered, rounded corners, minimum height=4em]
  \tikzstyle{line} = [draw, -latex', thick]
  \begin{tikzpicture}[auto]
      \node [block] (part1) {Part 1. For each $\mathbf{k}$-point, construct the overlap matrix $\mathbf{S}$.};
      \node [block, below=1cm of part1] (part2) {Part 2. Build $\mathbf{H}^{\alpha\alpha}$, $\mathbf{H}^{\alpha\beta}$, $\mathbf{H}^{\beta\alpha}$, and $\mathbf{H}^{\beta\beta}$, then assemble the noncollinear Hamiltonian $\mathbf{H}_{\mathrm{NC}}$.};
      \node [block, below=1cm of part2] (part3) {Part 3. Solve $\mathbf{H}_{\mathrm{NC}}\mathbf{C}_{\mathrm{NC}}=\mathbf{S}_{\mathrm{NC}}\mathbf{C}_{\mathrm{NC}}\mathbf{\varepsilon}$ for eigenvalues and eigenvectors.};
      \node [block, below=1cm of part3] (part4) {Part 4. Compute the chemical potential and the band energies.};
      \node [block, below=1cm of part4] (part5) {Part 5. Compute the charge-density and energy-density matrices.};
      \path[line] (part1) -- (part2);
      \path[line] (part2) -- (part3);
      \path[line] (part3) -- (part4);
      \path[line] (part4) -- (part5);
  \end{tikzpicture}
  \caption{Computational flow of the Band\_DFT\_NonCol routine.}
  \label{fig:flowchart_noncol}
\end{figure}
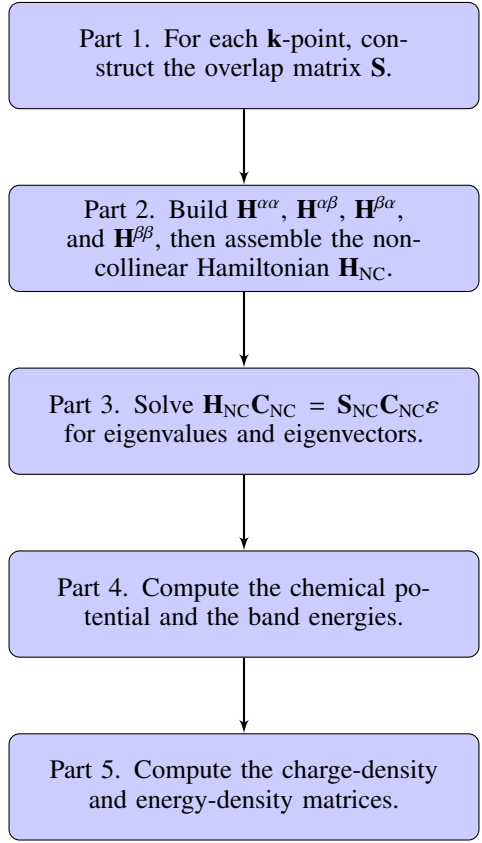

Figure~\ref{fig:flowchart_noncol} presents the computational flow of the Band\_DFT\_NonCol routine used for diagonalization in noncollinear DFT. Similar to the collinear routine, Parts 1 through 5 are collectively referred to as “diagonalization”.
Computations in Parts 1, 2, and 3 loop over assigned \(\mathbf{k}\)-points. The noncollinear Hamiltonian $\mathbf{H}_{\mathrm{NC}}$, coefficient matrix $\mathbf{C}_{\mathrm{NC}}$, and overlap matrix $\mathbf{S}_{\mathrm{NC}}$ are defined as:
\begin{equation}
    \mathbf{H}_{\mathrm{NC}}=\left(
        \begin{array}
        [c]{cc}%
        \mathbf{H}^{\mathrm{\alpha\alpha}} & \mathbf{H}^{\mathrm{\alpha\beta}}\\
        \mathbf{H}^{\mathrm{\beta\alpha}} & \mathbf{H}^{\mathrm{\beta\beta}}%
        \end{array}
        \right),
\end{equation}
\begin{equation}
    \mathbf{C}_{\mathrm{NC}}=\left(
        \begin{array}
        [c]{c}%
        \mathbf{C}^{\alpha}\\
        \mathbf{C}^{\beta}%
        \end{array}
        \right),
\end{equation}
\begin{equation}
    \mathbf{S}_{\mathrm{NC}}=\left(
        \begin{array}
        [c]{cc}%
        \mathbf{S} & \mathbf{0}\\
        \mathbf{0} & \mathbf{S}%
        \end{array}
        \right).
\end{equation}
The MPI parallelization strategy for $\mathbf{k}$-points mirrors the collinear case: when the number of MPI processes exceeds the number of $\mathbf{k}$-points, the loop over them is eliminated, and the computation for a single $\mathbf{k}$-point is distributed among multiple processes. Also consistent with the collinear implementation, Part 1 is MPI-parallelized in conventional OpenMX but executed serially in the GPU-accelerated version. GPU acceleration of Part 1 remains future work for the reasons discussed in Sect. 3.1.

\begin{figure}[bt]
  \centering
  \tikzstyle{block} = [rectangle, draw, fill=blue!20,
      text width=17em, text centered, rounded corners, minimum height=4em]
  \tikzstyle{line} = [draw, -latex', thick]
  \begin{tikzpicture}[auto]
      \node [block] (part31) {Part 3-1. Diagonalize $\mathbf{S}$ to obtain $\mathbf{U}$ and $\mathbf{s}$.};
      \node [block, below=1cm of part31] (part32) {Part 3-2. Construct $\mathbf{V}$ and $\mathbf{V}_{\mathrm{NC}}$ with $\mathbf{V}=\mathbf{U}\mathbf{s}^{-1/2}$.};
      \node [block, below=1cm of part32] (part33) {Part 3-3. Form $\mathbf{H}^{\prime \alpha\alpha}$, $\mathbf{H}^{\prime \alpha\beta}$, $\mathbf{H}^{\prime \beta\alpha}$, and $\mathbf{H}^{\prime \beta\beta}$ by similarity transformation.};
      \node [block, below=1cm of part33] (part34) {Part 3-4. Assemble $\mathbf{H}'_{\mathrm{NC}}$ from the transformed blocks.};
      \node [block, below=1cm of part34] (part35) {Part 3-5. Diagonalize $\mathbf{H}'_{\mathrm{NC}}$ to obtain eigenvalues and eigenvectors $\mathbf{C}'_{\mathrm{NC}}$.};
      \node [block, below=1cm of part35] (part36) {Part 3-6. Recover $\mathbf{C}_{\mathrm{NC}}=\mathbf{V}_{\mathrm{NC}}\mathbf{C}'_{\mathrm{NC}}$.};
      \path[line] (part31) -- (part32);
      \path[line] (part32) -- (part33);
      \path[line] (part33) -- (part34);
      \path[line] (part34) -- (part35);
      \path[line] (part35) -- (part36);
  \end{tikzpicture}
  \caption{Detailed flow of Part 3 in the Band\_DFT\_NonCol routine.}
  \label{fig:flowchart_noncol_part3}
\end{figure}
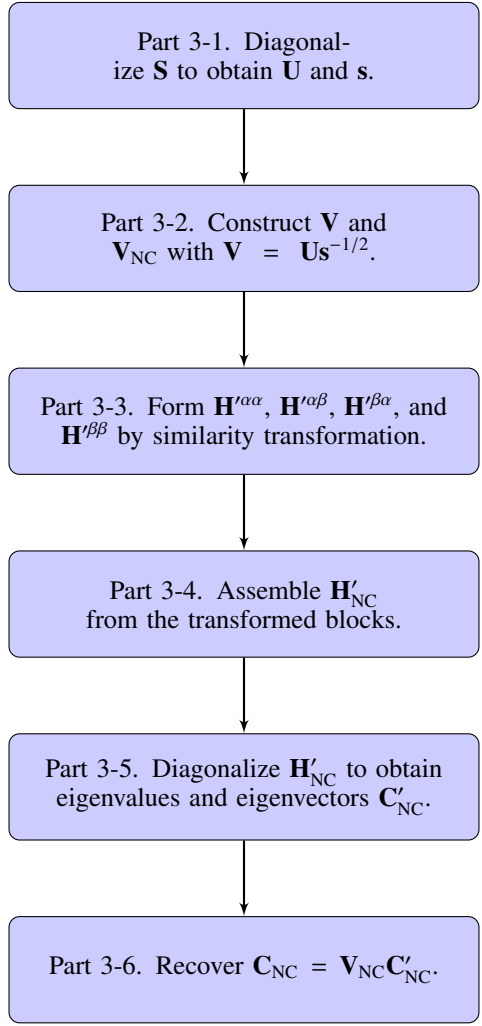

The procedure in Part 3 of the Band\_DFT\_NonCol routine is subdivided as shown in Fig.~\ref{fig:flowchart_noncol_part3}.

In Part 3-1, $\mathbf{U}$ is the unitary matrix that diagonalizes $\mathbf{S}$, and $\mathbf{s}$ is the diagonalized form of $\mathbf{S}$. We define:
\begin{equation}
    \mathbf{V}_{\mathrm{NC}}=\left(
        \begin{array}
        [c]{cc}%
        \mathbf{V} & \mathbf{0}\\
        \mathbf{0} & \mathbf{V}%
        \end{array}
        \right),
\end{equation}
\begin{equation}
    \mathbf{H}_{\mathrm{NC}}^{\prime}=\left(
        \begin{array}
        [c]{cc}%
        \mathbf{H}^{\prime \mathrm{\alpha\alpha}} & \mathbf{H}^{\prime \mathrm{\alpha\beta}}\\
        \mathbf{H}^{\prime \mathrm{\beta\alpha}} & \mathbf{H}^{\prime \mathrm{\beta\beta}}%
        \end{array}
        \right),
\end{equation}
\begin{equation}
    \mathbf{C}_{\mathrm{NC}}^{\prime}=\left(
        \begin{array}
        [c]{c}%
        \mathbf{C}^{\prime\alpha}\\
        \mathbf{C}^{\prime\beta}%
        \end{array}
        \right).
\end{equation}
The matrix dimensions of $\mathbf{V}_{\mathrm{NC}}$ and $\mathbf{H}'_{\mathrm{NC}}$ are twice the total number of basis functions, whereas all other matrices have dimensions equal to the total number of basis functions.

Similar to the collinear approach, we disable MPI parallelization for Parts 1 through 3-6, assigning the full-size matrix to a single MPI process. We replace ELPA with cuSOLVER in Parts 3-1 and 3-5, and ScaLAPACK with cuBLAS in Parts 3-3 and 3-6. Additionally, OpenACC is used to accelerate portions of Part 2 and Part 3-2, and all of Part 3-4. As in the collinear case, the lack of Tensor Core usage in these auxiliary steps has a negligible impact on overall performance.

Procedures from Part 4 onward require block-cyclically distributed matrices. Therefore, we redistribute the full-size matrix into a block-cyclic layout at the end of Part 3-6. We do not implement GPU acceleration for Part 4 because its computational cost is negligible (see Table~\ref{tab:breakdown_ncdft_gpu}). GPU acceleration of Part 5 is challenging for similar reasons to Part 4 in the collinear case. Therefore, Parts 4 and 5 remain MPI-parallelized using the original OpenMX code. These changes require modifications to the Band\_DFT\_NonCol routine.

\section{Calculation and Measurement Conditions}
We perform benchmark calculations with OpenMX version 3.9.9, measuring execution time in two settings: (i) CPU-only and (ii) GPU-accelerated (utilizing both CPU and GPU). Benchmarks are run on the Pegasus supercomputer at the University of Tsukuba. Each node contains one 48-core Intel Xeon Platinum 8468 CPU and one Hopper-generation NVIDIA H100 GPU. Both configurations use the NVIDIA HPC Compiler 23.1, CUDA 12.0, Intel oneMKL 2023.0, and Open MPI 4.1.8. For the underlying parallel linear algebra, we use the ScaLAPACK library provided with oneMKL. For the eigensolver, we employ the two-stage method of the modified ELPA version 2018.05.001 bundled with OpenMX. 
cuBLAS and cuSOLVER are the versions shipped with CUDA 12.0. 

The benchmarks use supercells of diamond-structured silicon ($a = 5.431\,\text{\AA}$)~\cite{crystal}, generated with OpenMX Viewer~\cite{openmx_viewer}. All calculations employ the GGA-PBE exchange--correlation functional~\cite{pbe}, a $2\times2\times2$ $\mathbf{k}$-point mesh, the \textit{s}2\textit{p}2\textit{d}1 standard basis set (13 basis functions per atom), and an energy cutoff of 200 Ry. The number of SCF iterations is fixed at 25 for collinear DFT and 22 for noncollinear DFT. All benchmarks use a number of MPI processes equal to the total number of CPU cores (48 per node). Given the $2\times2\times2$ mesh, the number of irreducible $\mathbf{k}$-points calculated is 4 for collinear DFT and 8 for noncollinear DFT.

Because cuBLAS and cuSOLVER process the full-size matrix for each $\mathbf{k}$-point on a single GPU, the number of utilized GPUs cannot exceed the number of $\mathbf{k}$-points being computed; excess GPUs remain idle. Conversely, a single GPU can process multiple $\mathbf{k}$-points (associated with multiple MPI processes). Therefore, we adopt 4 nodes (4 GPUs) for collinear DFT and 8 nodes (8 GPUs) for noncollinear DFT as baseline configurations. Due to host (CPU) memory limitations, the collinear benchmarks are limited to 1,200 atoms (on 4 nodes), and the noncollinear benchmarks are limited to 640 atoms (on 8 nodes). OpenMP thread parallelism is not enabled in this study.

We enable NVIDIA Multi-Process Service (MPS) when multiple MPI processes share a single GPU, although it does not yield a performance gain in these benchmarks. Load-balance data, GPU utilization, and host-device (GPU) transfer times are measured using NVIDIA Nsight Systems, taking into account the overlap between kernel execution and host-device data transfers. Each benchmark is executed three times, and the average is reported. In all cases, the absolute difference in total energy between CPU-only and GPU-accelerated calculations is below $10^{-8}$ Hartree.

\section{GPU Acceleration Results}
\subsection{Results for collinear DFT}
We first summarize the breakdown of the collinear-DFT wall time for a 1,200-atom silicon system (basis-set size 15,600) computed on four CPU-only nodes (192 CPU cores) in Table~\ref{tab:dft_cpu_time}.
\begin{table}[bt]
  \centering
  \caption{Breakdown of collinear-DFT wall time for CPU-only calculation. The measurements are performed on a 1,200-atom supercell of diamond-structure silicon.}
  \label{tab:dft_cpu_time}
  \begin{tabular}{c c c}
  \hline
  Routine & Time (sec) & Proportion (\%) \\
  \hline
  Set OLP Kin & 3.595 & 0.15 \\
  Set Nonlocal & 5.694 & 0.24 \\
  Set ProExpn VNA & 17.51 & 0.73 \\
  Set Hamiltonian & 75.77 & 3.14 \\
  Diagonalization & 2268 & 94.1 \\
  Mixing DM & 1.187 & 0.05 \\
  Force & 10.25 & 0.42 \\
  Total Energy & 6.364 & 0.26 \\
  Set Density Grid & 16.88 & 0.70 \\
  FFT(2D) Density & 1.3219 & 0.05 \\
  Others & 4.311 & 0.18 \\
  Total & 2411 & 100.0 \\
  \hline
  \end{tabular}
\end{table}
The diagonalization step dominates the execution, accounting for 94.1\% of the total wall time.
Table~\ref{tab:breakdown_scf}(a) details the diagonalization runtime (maximum across 192 MPI processes) for the first SCF cycle under the same conditions. The eigenvalue solvers (Parts 2-1 and 2-4) account for 78.2\% of the time, while matrix multiplications (Parts 2-3 and 2-5) constitute 19.3\%.

Table~\ref{tab:breakdown_scf}(b) shows the breakdown for the subsequent SCF cycles (average of cycles 2--25, maximum across MPI processes). Note that due to this per-cycle analysis and averaging, the total times in Table~\ref{tab:breakdown_scf} do not directly correspond to the total “Diagonalization” time in Table~\ref{tab:dft_cpu_time}. From the second cycle onward, matrices $\mathbf{S}$ and $\mathbf{V}$ are reused, eliminating the need for Parts 2-1 and 2-2. In these cycles, the eigenvalue solver (Part 2-4) accounts for 52.0\% of the time, and matrix multiplications (Parts 2-3 and 2-5) contribute 43.7\%.

\begin{table}[bt]
  \centering
  \caption{Detailed diagonalization timing for collinear DFT (CPU-only). (a) First SCF cycle. (b) SCF cycles 2--25 (average). Measurements are conducted on a 1,200-atom supercell of diamond-structure silicon.}
  \label{tab:breakdown_scf}
  \begin{minipage}[t]{0.48\textwidth}
    \centering
    \begin{tabular}[t]{c c c}
      \toprule
      \begingroup
  \setbox0=\hbox{\shortstack{Routine}}%
  \setbox1=\hbox{\footnotesize(a)}%
  \makebox[\dimexpr \wd1 + 0.60em + \wd0\relax][l]{%
    \raisebox{\dimexpr (\ht0+\dp0-\ht1-\dp1)/2 + 0.10ex\relax}{\copy1}%
    \hspace{0.60em}%
    \copy0
  }%
\endgroup & Time (sec) & Proportion (\%) \\
      \midrule
      Part 1 & 1.654 & 0.82 \\
      part 2-1 & 112.0 & 55.8 \\
      part 2-2 & 0.1422 & 0.07 \\
      part 2-3 & 24.30 & 12.1 \\
      part 2-4 & 44.91 & 22.4 \\
      part 2-5 & 14.55 & 7.24 \\
      Part 3 & 0.0107 & 0.01 \\
      Part 4 & 3.285 & 1.64 \\
      Total & 200.9 & 100.0 \\
      \bottomrule
    \end{tabular}
  \end{minipage}\hfill
  \begin{minipage}[t]{0.48\textwidth}
    \centering
    \begin{tabular}[t]{c c c}
      \toprule
      \begingroup
  \setbox0=\hbox{\shortstack{Routine}}%
  \setbox1=\hbox{\footnotesize(b)}%
  \makebox[\dimexpr \wd1 + 0.60em + \wd0\relax][l]{%
    \raisebox{\dimexpr (\ht0+\dp0-\ht1-\dp1)/2 + 0.10ex\relax}{\copy1}%
    \hspace{0.60em}%
    \copy0
  }%
\endgroup & Time (sec) & Proportion (\%) \\
      \midrule
      Part 1 & 0.6689 & 0.77 \\
      part 2-1 & 0.0000 & 0.00 \\
      part 2-2 & 0.0000 & 0.00 \\
      part 2-3 & 24.29 & 28.1 \\
      part 2-4 & 44.98 & 52.0 \\
      part 2-5 & 13.49 & 15.6 \\
      Part 3 & 0.0097 & 0.01 \\
      Part 4 & 3.106 & 3.59 \\
      Total & 86.54 & 100.0 \\
      \bottomrule
    \end{tabular}
  \end{minipage}
\end{table}
Table~\ref{tab:dft_cpu_time_gpu} reports the wall-time breakdown when four GPUs are added to the configuration (four nodes, four GPUs). Even with GPU acceleration, diagonalization remains the dominant cost at 82.5\% of the total wall time.

\begin{table}[bt]
  \centering
  \caption{Breakdown of collinear-DFT wall time with GPU acceleration. Measurements are conducted on a 1,200-atom supercell of diamond-structure silicon.}
  \label{tab:dft_cpu_time_gpu}
  \begin{tabular}{c c c}
  \hline
  Routine & Time (sec) & Proportion (\%) \\
  \hline
  Set OLP Kin & 3.605 & 0.43 \\
  Set Nonlocal & 5.761 & 0.68 \\
  Set ProExpn VNA & 17.36 & 2.06 \\
  Set Hamiltonian & 80.23 & 9.53 \\
  Diagonalization & 694.2 & 82.5 \\
  Mixing DM & 1.334 & 0.16 \\
  Force & 10.23 & 1.21 \\
  Total Energy & 5.649 & 0.67 \\
  Set Density Grid & 17.11 & 2.03 \\
  FFT(2D) Density & 1.526 & 0.18 \\
  Others & 4.689 & 0.56 \\
  Total & 841.7 & 100.0 \\
  \hline
  \end{tabular}
\end{table}

Table~\ref{tab:breakdown_scf_gpu}(a) details the diagonalization time for the first SCF iteration with GPU acceleration (maximum across 192 MPI processes). The eigenvalue problems (Parts 2-1 and 2-4) represent 44.3\% of the time, and matrix multiplications (Parts 2-3 and 2-5) contribute 18.2\%. The combined share of these four steps decreases significantly from 97.5\% in the CPU-only case [Table~\ref{tab:breakdown_scf}(a)] to 62.5\% with GPU acceleration.
Notably, the elapsed time for Part 1 increases substantially compared to the CPU-only calculation. This is because Part 1 must be executed serially (no MPI parallelization) in the GPU implementation, as discussed in Sect. 3.1. 

Furthermore, Part 2-5 includes the overhead of redistributing the full matrices from a single MPI process into a block-cyclic layout, accounting for 5.26\% of the diagonalization time.
Relative to the CPU-only calculation, the eigenvalue solutions accelerate by 8.15 times (Part 2-1) and 4.77 times (Part 2-4), while the matrix products in Part 2-3 accelerate by 11.1 times.

Table~\ref{tab:breakdown_scf_gpu}(b) shows the breakdown for subsequent SCF cycles (average of cycles 2--25). The eigenvalue problem (Part 2-4) accounts for 36.0\% of the diagonalization time, and matrix multiplications (Parts 2-3 and 2-5) account for 26.7\%.

The combined share of Parts 2-3, 2-4, and 2-5 reduces from 95.7\% (CPU-only) to 62.7\% (GPU-accelerated). The speedups achieved are 8.82 times for Parts 2-3 and 4.74 times for Part 2-4. The apparent reduction in speedup for Part 2-3 compared to the first iteration (11.1 times) is due to how host-device data transfer times are measured. In the first SCF cycle, transfer times are partially included in both Parts 2-1 and 2-3. From the second cycle onward (where Part 2-1 is skipped), the full transfer overhead is attributed to Part 2-3, lowering its apparent speedup. Furthermore, the overhead for distributing matrices accounts for 8.92\% of the diagonalization time.
\begin{table}[bt]
  \centering
  \caption{Breakdown of diagonalization time for collinear DFT with GPU acceleration. (a) First SCF cycle. (b) SCF cycles 2--25 (average). Measurements are conducted on a 1,200-atom supercell of diamond-structure silicon.}
  \label{tab:breakdown_scf_gpu}
  \begin{minipage}[t]{0.48\textwidth}
    \centering
    \begin{tabular}[t]{c c c}
      \toprule
      \begingroup
  \setbox0=\hbox{\shortstack{Routine}}%
  \setbox1=\hbox{\footnotesize(a)}%
  \makebox[\dimexpr \wd1 + 0.60em + \wd0\relax][l]{%
    \raisebox{\dimexpr (\ht0+\dp0-\ht1-\dp1)/2 + 0.10ex\relax}{\copy1}%
    \hspace{0.60em}%
    \copy0
  }%
\endgroup & Time (sec) & Proportion (\%) \\
      \midrule
      Part 1 & 15.68 & 30.0 \\
      part 2-1 & 13.74 & 26.3 \\
      part 2-2 & 0.5939 & 1.14 \\
      part 2-3 & 2.185 & 4.18 \\
      part 2-4 & 9.415 & 18.0 \\
      part 2-5 & 7.306 & 14.0 \\
      (MPI) & (2.749) & (5.26) \\
      Part 3 & 0.0173 & 0.03 \\
      Part 4 & 3.301 & 6.32 \\
      Total & 52.23 & 100.0 \\
      \bottomrule
    \end{tabular}
  \end{minipage}\hfill
  \begin{minipage}[t]{0.48\textwidth}
    \centering
    \begin{tabular}[t]{c c c}
      \toprule
      \begingroup
  \setbox0=\hbox{\shortstack{Routine}}%
  \setbox1=\hbox{\footnotesize(b)}%
  \makebox[\dimexpr \wd1 + 0.60em + \wd0\relax][l]{%
    \raisebox{\dimexpr (\ht0+\dp0-\ht1-\dp1)/2 + 0.10ex\relax}{\copy1}%
    \hspace{0.60em}%
    \copy0
  }%
\endgroup & Time (sec) & Proportion (\%) \\
      \midrule
      Part 1 & 6.703 & 25.5 \\
      part 2-1 & 0.0000 & 0.00 \\
      part 2-2 & 0.0000 & 0.00 \\
      part 2-3 & 2.755 & 10.5 \\
      part 2-4 & 9.481 & 36.0 \\
      part 2-5 & 4.262 & 16.2 \\
      (MPI) & 2.347 & 8.92 \\
      Part 3 & 0.0157 & 0.06 \\
      Part 4 & 3.109 & 11.8 \\
      Total & 26.32 & 100.0 \\
      \bottomrule
    \end{tabular}
  \end{minipage}
\end{table}

\begin{figure}[bt]
  \centering
  \includegraphics[width=0.49\textwidth]{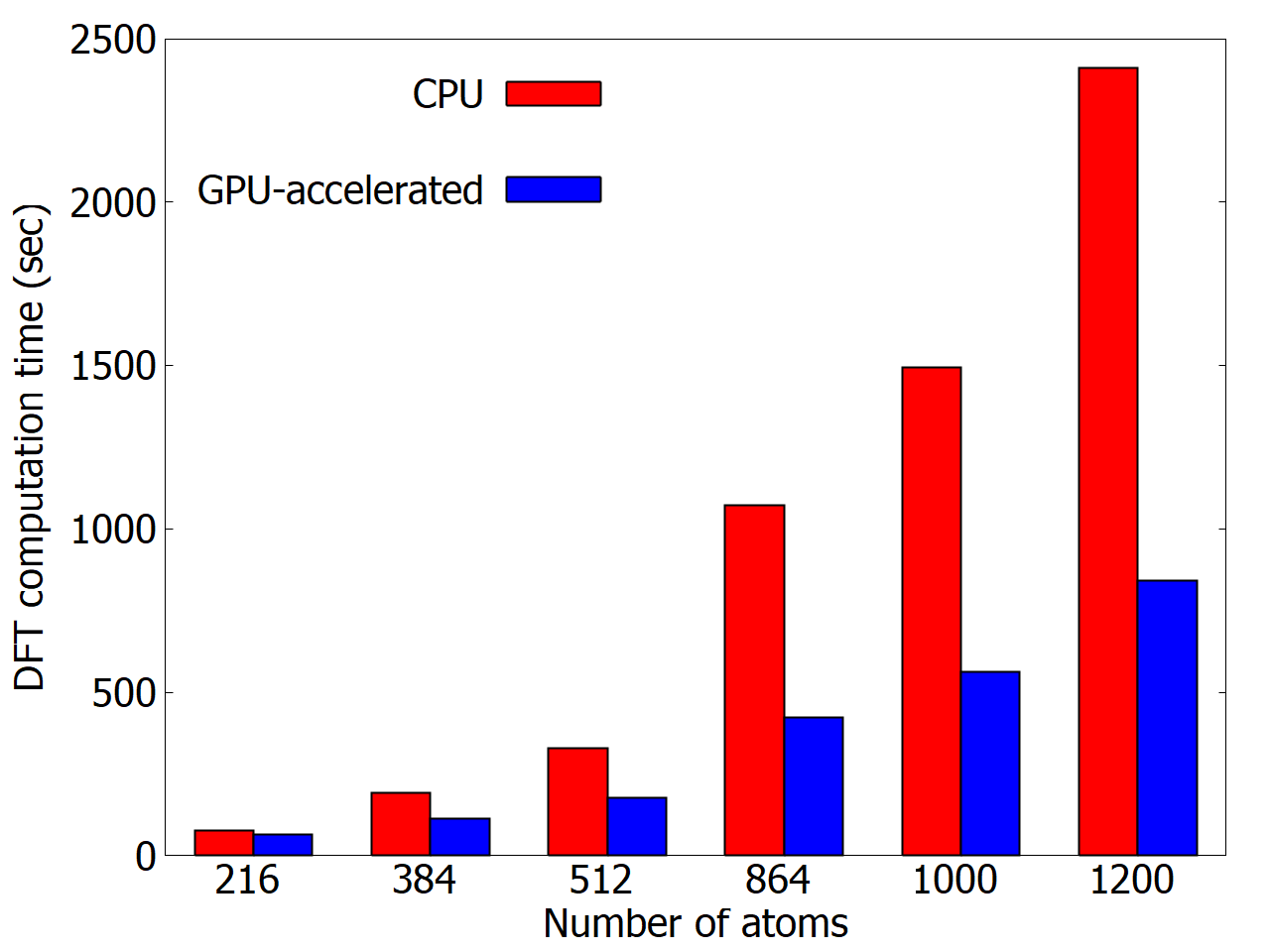}
  \caption{Relationship between number of atoms and total DFT wall time for collinear DFT (Four nodes).}
  \label{fig:atoms_bar_graph}
\end{figure}

Figure~\ref{fig:atoms_bar_graph} and Table~\ref{tab:atoms} compare the total wall time versus number of atoms on four nodes for CPU-only and GPU-accelerated calculations. The GPU implementation outperforms the CPU implementation for all system sizes, reaching a maximum speedup of 2.86 times at 1,200 atoms.
\begin{table}[bt]
  \centering
  \caption{Variation of total DFT wall time (s) with number of atoms for CPU and GPU-accelerated collinear DFT (Four nodes).}
  \label{tab:atoms}
  \begin{tabular}{c c c c}
  \hline
  \rule{0pt}{5ex}
  \shortstack{Number\\of atoms} & CPU & \shortstack{GPU-\\accelerated} & \shortstack{Speed-up\\(times)} \\
  \hline
  216  &   77.8 &  66.4 & 1.17 \\
  384  &  191.6 & 113.7 & 1.69 \\
  512  &  328.6 & 176.2 & 1.86 \\
  864  & 1073.6 & 424.4 & 2.53 \\
  1000 & 1494.3 & 563.7 & 2.65 \\
  1200 & 2411.0 & 841.7 & 2.86 \\
  \hline
  \end{tabular}
\end{table}

\begin{figure}[bt]
  \centering
  \includegraphics[width=0.49\textwidth]{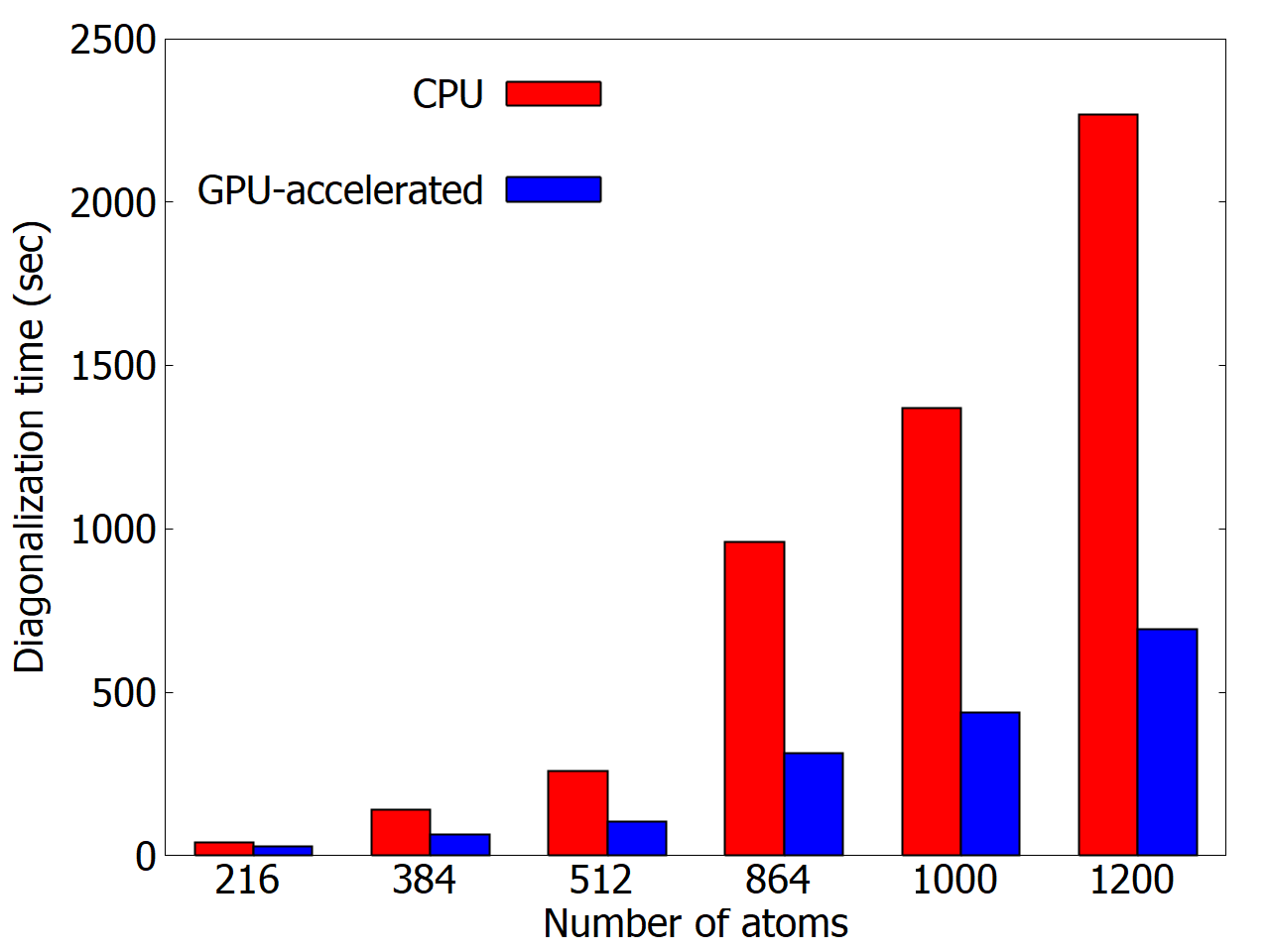}
  \caption{Relationship between number of atoms and diagonalization time for collinear DFT (Four nodes).}
  \label{fig:atoms_diag_bar_graph}
\end{figure}

Figure~\ref{fig:atoms_diag_bar_graph} and Table~\ref{tab:atoms_diag} compare the corresponding diagonalization times. The speedup reaches 3.27 times at 1,200 atoms. This result is discussed further in Sect. 6.

\begin{table}[bt]
  \centering
  \caption{Variation of diagonalization time (s) with number of atoms for CPU and GPU-accelerated collinear DFT (Four nodes).}
  \label{tab:atoms_diag}
  \begin{tabular}{c c c c}
  \hline
  \rule{0pt}{5ex}
  \shortstack{Number\\of atoms} & CPU & \shortstack{GPU-\\accelerated} & \shortstack{Speed-up\\(times)} \\
  \hline
  216  &   41.0 &  28.1 & 1.46 \\
  384  &  140.8 &  65.1 & 2.16 \\
  512  &  260.2 & 104.0 & 2.50 \\
  864  &  960.4 & 313.2 & 3.07 \\
  1000 & 1370.3 & 437.0 & 3.14 \\
  1200 & 2268.1 & 694.2 & 3.27 \\
  \hline
  \end{tabular}
\end{table}

\begin{figure}[bt]
    \centering
    \begin{minipage}[t]{0.48\textwidth}
      \centering
      \includegraphics[width=\linewidth]{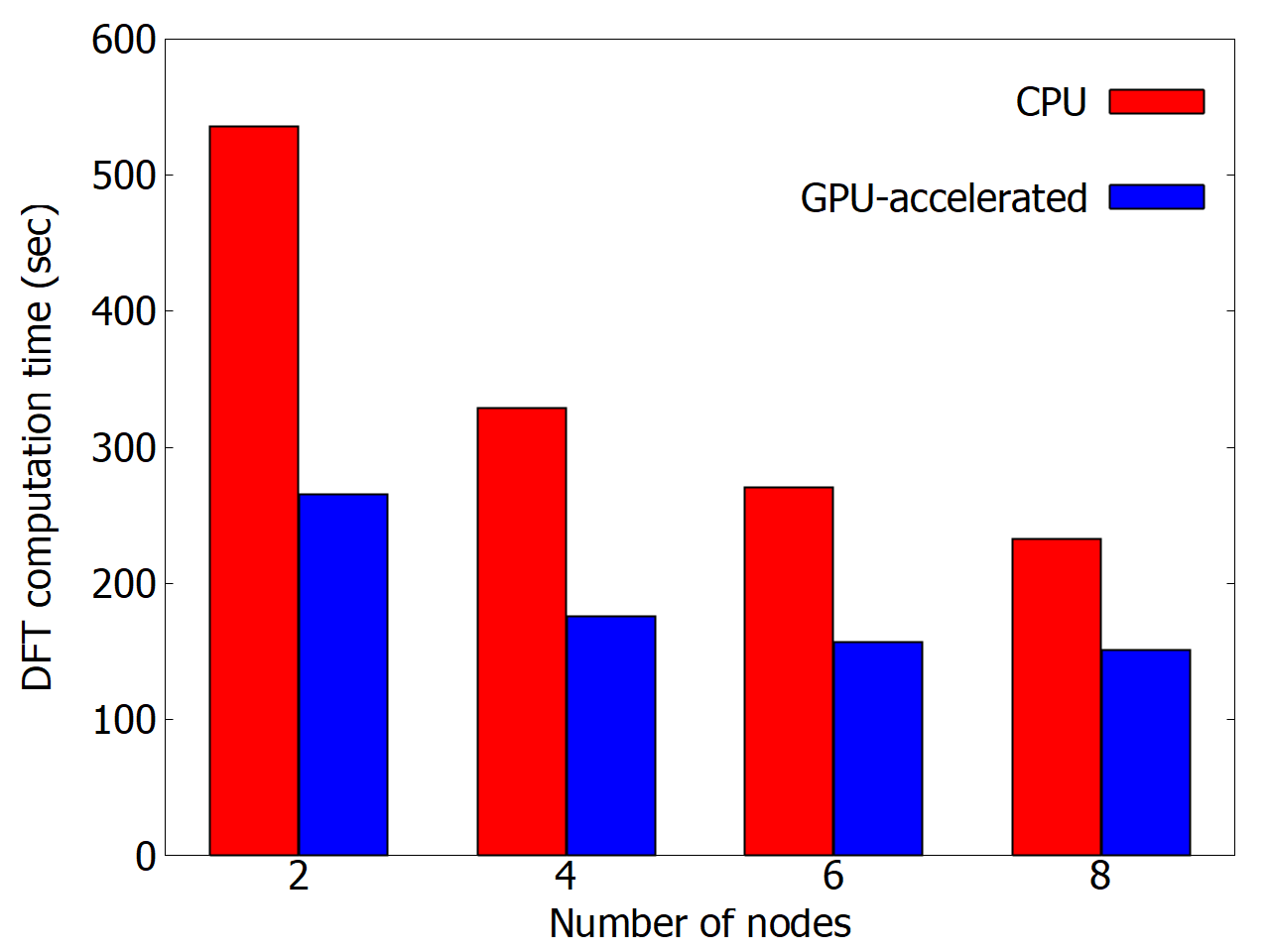}
      \footnotesize{(a)}
    \end{minipage}
    \hfill
    \begin{minipage}[t]{0.48\textwidth}
      \centering
      \includegraphics[width=\linewidth]{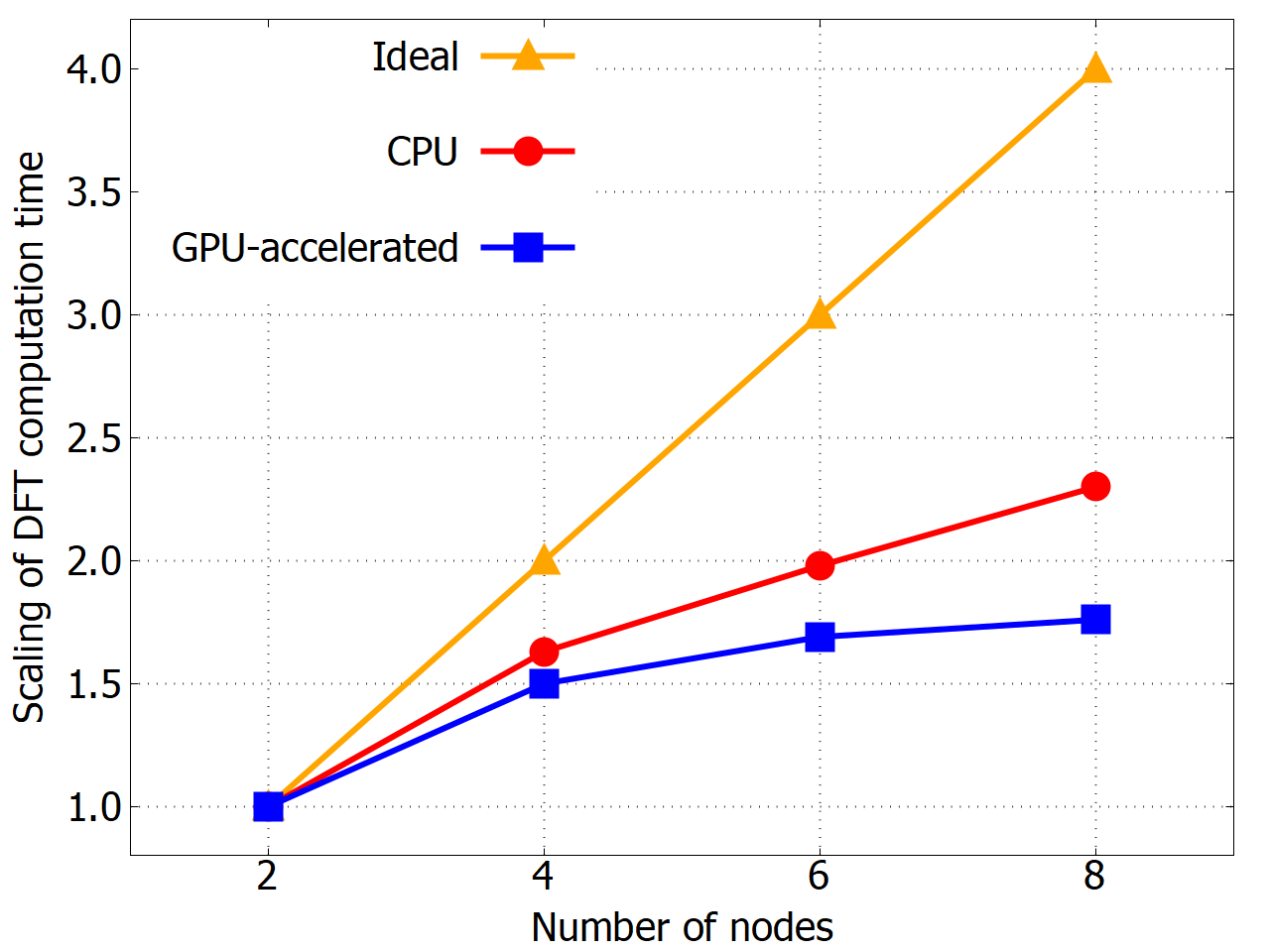}
      \footnotesize{(b)}
    \end{minipage}
    \caption{Speed comparison between CPU-only and GPU-accelerated collinear DFT as the number of nodes increases. (a) Comparison of total DFT wall time. (b) Scaling comparison (relative to 2 nodes). Measurements are performed on a 512-atom supercell of diamond-structure silicon.}
    \label{fig:nodes}
\end{figure}

Figure~\ref{fig:nodes}(a) compares the DFT wall times for a fixed system size (512 atoms) as the node count increases from 2 to 8. Figure~\ref{fig:nodes}(b) presents the corresponding scaling, defined as the speedup relative to the two-node wall time. Table~\ref{tab:nodes} provides the numerical data, including speedup factors and scaling behavior.
The scaling behavior of the GPU-accelerated implementation is poor [Fig.~\ref{fig:nodes}(b), Table~\ref{tab:nodes}(b)]. The GPU acceleration factor decreases as the node count increases: the calculation is 2.02 times faster than the CPU-only calculation on two nodes, but only 1.54 times faster on eight nodes.

\begin{table}[bt]
  \centering
  \caption{Variation of total DFT wall time (s) and scaling with node count for CPU and GPU-accelerated collinear DFT. (a) Total DFT wall time. (b) Scaling factor. The system is the same as for Fig.~\ref{fig:nodes}.}
  \label{tab:nodes}
  \begin{minipage}[t]{0.48\textwidth}
    \centering
    \begin{tabular}[t]{c c c c}
  \hline
  \noalign{\vskip 0.6ex}
  \begingroup
  \setbox0=\hbox{\shortstack{Number\\ of nodes}}%
  \setbox1=\hbox{\footnotesize(a)}%
  \makebox[\dimexpr \wd1 + 0.60em + \wd0\relax][l]{%
    \raisebox{\dimexpr (\ht0+\dp0-\ht1-\dp1)/2 + 0.10ex\relax}{\copy1}%
    \hspace{0.60em}%
    \copy0
  }%
\endgroup & CPU &
  \shortstack{GPU-\\accelerated} &
  \shortstack{Speed-up\\(times)}
  \\
  \hline
  2 & 535.7 & 265.2 & 2.02 \\
  4 & 328.6 & 176.2 & 1.86 \\
  6 & 270.8 & 157.0 & 1.72 \\
  8 & 232.8 & 150.8 & 1.54 \\
  \hline
\end{tabular}
  \end{minipage}
  \hfill
  \begin{minipage}[t]{0.48\textwidth}
    \centering
    \begin{tabular}[t]{c c c}
    \hline
    \noalign{\vskip 0.6ex}
    \begingroup
  \setbox0=\hbox{\shortstack{Number\\ of nodes}}%
  \setbox1=\hbox{\footnotesize(b)}%
  \makebox[\dimexpr \wd1 + 0.60em + \wd0\relax][l]{%
    \raisebox{\dimexpr (\ht0+\dp0-\ht1-\dp1)/2 + 0.10ex\relax}{\copy1}%
    \hspace{0.60em}%
    \copy0
  }%
\endgroup & CPU Scaling & GPU Scaling \\
    \hline
    2 & 1.00 & 1.00 \\
    4 & 1.63 & 1.50 \\
    6 & 1.98 & 1.69 \\
    8 & 2.30 & 1.76 \\
    \hline
    \end{tabular}
  \end{minipage}
\end{table}

\begin{figure}[bt]
  \centering
  \begin{minipage}[t]{0.48\textwidth}
    \centering
    \includegraphics[width=\linewidth]{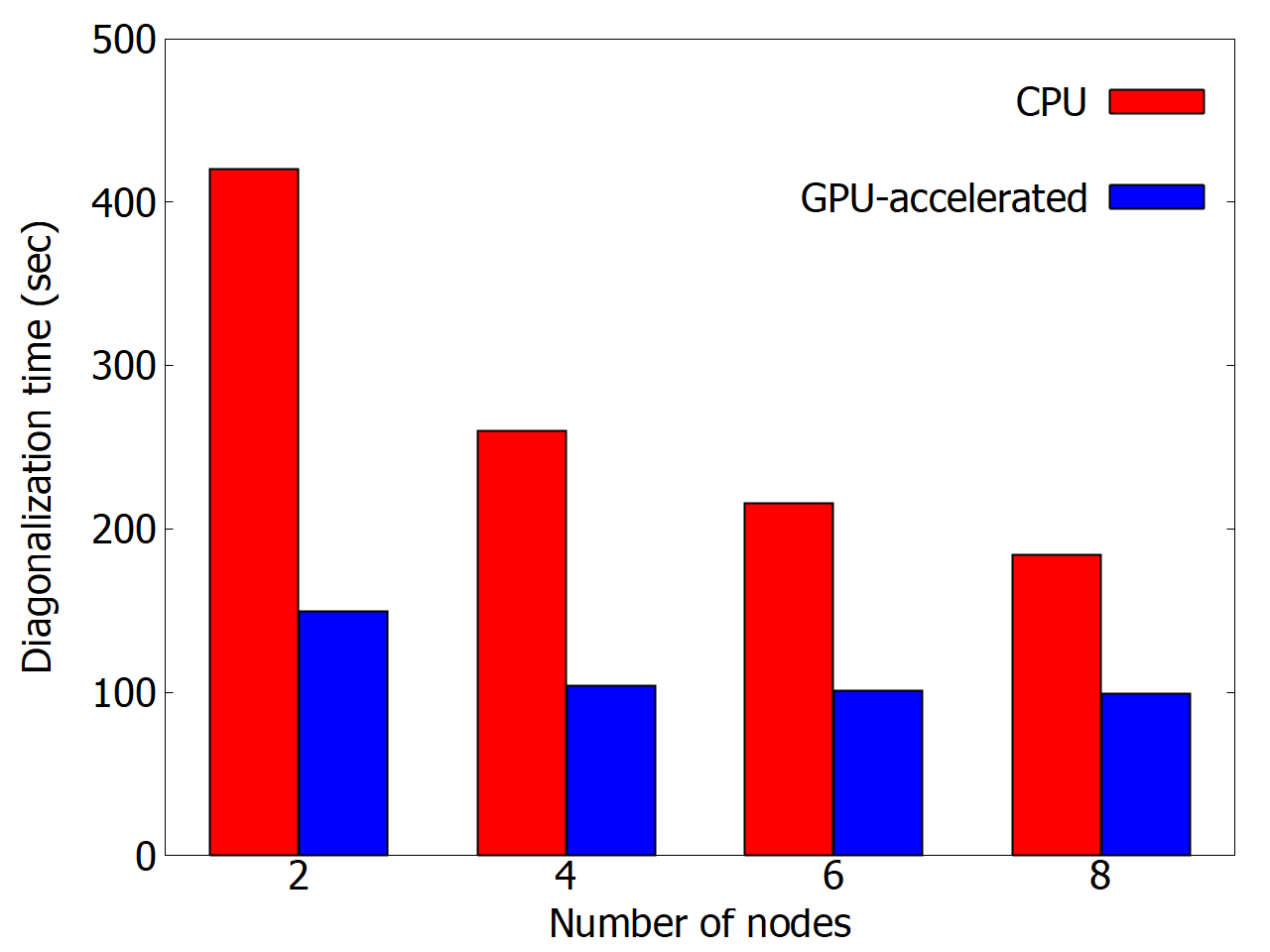}
    \footnotesize{(a)}
  \end{minipage}
  \hfill
  \begin{minipage}[t]{0.48\textwidth}
    \centering
    \includegraphics[width=\linewidth]{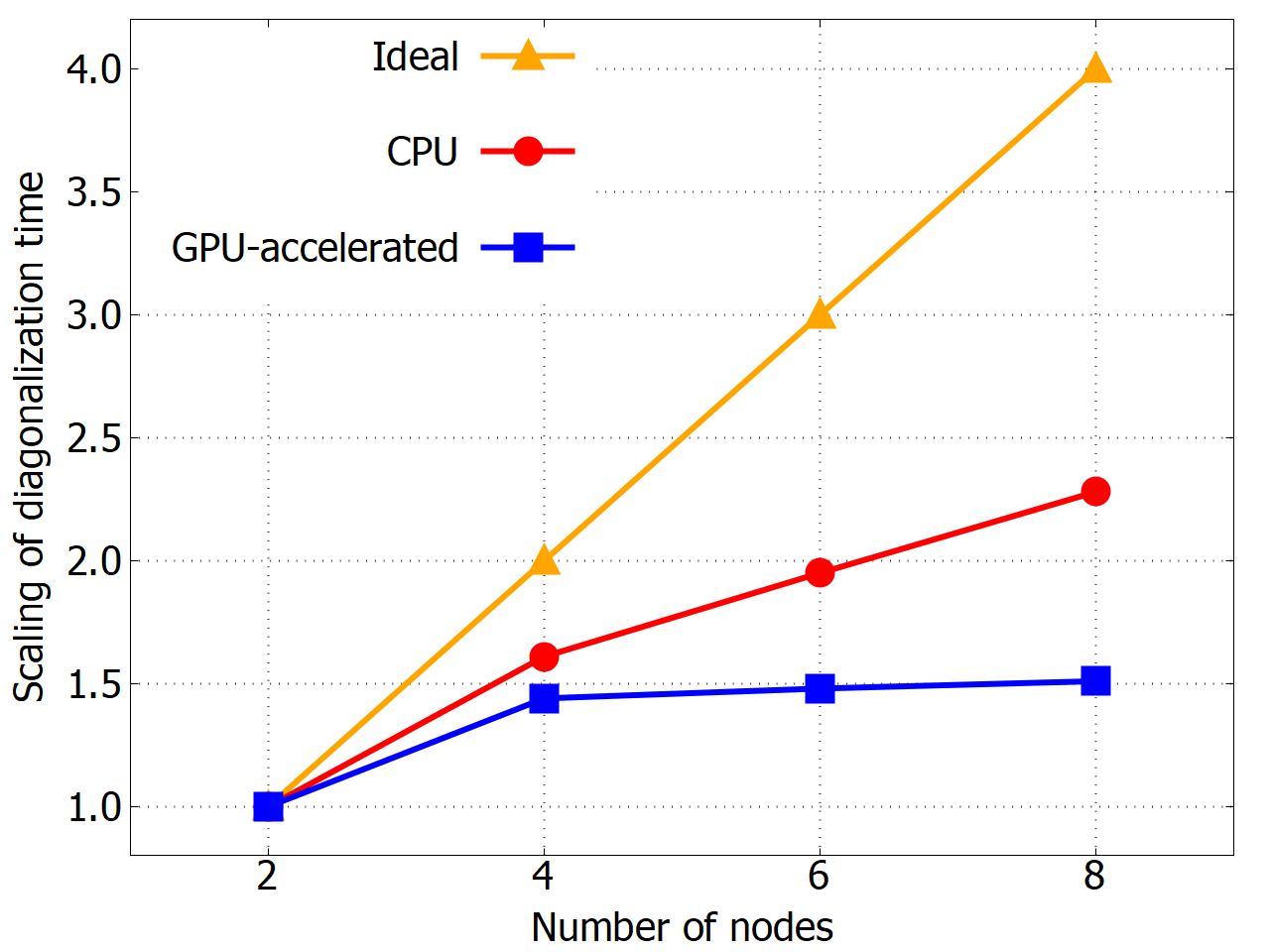}
    \footnotesize{(b)}
  \end{minipage}
  \caption{Speed comparison for diagonalization between CPU-only and GPU-accelerated collinear DFT as the number of nodes increases. (a) Comparison of diagonalization time. (b) Scaling comparison. The system is the same as for Fig.~\ref{fig:nodes}.}
  \label{fig:nodes_diag}
\end{figure}
\begin{table}[bt]
  \centering
  \caption{Variation of diagonalization time (s) and scaling with node count for CPU and GPU-accelerated collinear DFT. (a) Diagonalization time. (b) Scaling. The system is the same as for Fig.~\ref{fig:nodes}.}
  \label{tab:nodes_diag}
  \begin{minipage}[t]{0.48\textwidth}
    \centering
    \begin{tabular}[t]{c c c c}
    \hline
    \rule{0pt}{5ex}
    \begingroup
  \setbox0=\hbox{\shortstack{Number\\ of nodes}}%
  \setbox1=\hbox{\footnotesize(a)}%
  \makebox[\dimexpr \wd1 + 0.60em + \wd0\relax][l]{%
    \raisebox{\dimexpr (\ht0+\dp0-\ht1-\dp1)/2 + 0.10ex\relax}{\copy1}%
    \hspace{0.60em}%
    \copy0
  }%
\endgroup & CPU & \shortstack{GPU-\\accelerated} & \shortstack{Speed-up\\(times)} \\
    \hline
    2 & 420.2 & 149.5 & 2.81 \\
    4 & 260.2 & 104.0 & 2.50 \\
    6 & 215.8 & 101.3 & 2.13 \\
    8 & 184.0 &  99.3 & 1.85 \\
    \hline
    \end{tabular}
  \end{minipage}
  \hfill
  \begin{minipage}[t]{0.48\textwidth}
    \centering
    \begin{tabular}[t]{c c c}
    \hline
    \noalign{\vskip 0.6ex}
    \begingroup
  \setbox0=\hbox{\shortstack{Number\\ of nodes}}%
  \setbox1=\hbox{\footnotesize(b)}%
  \makebox[\dimexpr \wd1 + 0.60em + \wd0\relax][l]{%
    \raisebox{\dimexpr (\ht0+\dp0-\ht1-\dp1)/2 + 0.10ex\relax}{\copy1}%
    \hspace{0.60em}%
    \copy0
  }%
\endgroup & CPU Scaling & GPU Scaling \\
    \hline
    2 & 1.00 & 1.00 \\
    4 & 1.61 & 1.44 \\
    6 & 1.95 & 1.48 \\
    8 & 2.28 & 1.51 \\
    \hline
    \end{tabular}
  \end{minipage}
\end{table}
The CPU-only implementation (MPI-parallelized via ScaLAPACK and ELPA) exhibits more favorable scaling than its GPU-accelerated counterpart, achieving a 2.30 times speedup when increasing from 2 to 8 nodes [Table \ref{tab:nodes}(b)]. We note that the elapsed times for ScaLAPACK and ELPA include MPI communication overhead, which is difficult to isolate and measure directly.
Figure~\ref{fig:nodes_diag} shows the corresponding diagonalization times (a) and scaling behavior (b) for the 512-atom system.

Similar to the total wall time, the scaling of the diagonalization step is poor with GPU acceleration [Fig.~\ref{fig:nodes_diag}(b), Table~\ref{tab:nodes_diag}(b)]. The GPU speedup decreases from 2.81 times on two nodes to 1.85 times on eight nodes. Performance saturates at six nodes and above because only four $\mathbf{k}$-points are calculated, meaning only four GPUs can be utilized. The limited performance gain (1.44 times) when moving from two to four GPUs is analyzed in Sect. 6. In contrast, the CPU-only configuration shows better scaling, achieving a 2.28 times speedup from 2 to 8 nodes.

All subsequent results are reported as per-GPU averages.

We assess GPU load balance using the coefficient of variation for GPU execution times (including kernel time only, or kernel plus data transfer time). A coefficient near 0\% indicates perfect balance.

Table~\ref{tab:load_balance}(a) presents the load balance as the number of atoms increases (four nodes fixed). The 1,200-atom system is omitted due to memory exhaustion during measurement. In all cases, the coefficient remains below 1\%, indicating excellent load balance.
Table~\ref{tab:load_balance}(b) shows the load balance when varying the node count (512 atoms fixed). Cases with six or more nodes are excluded as only four GPUs are utilized. The coefficient stays under 1\%, again demonstrating excellent load balance.
\begin{table}[bt]
  \centering
  \caption{GPU load balance (coefficient of variation) for collinear DFT. (a) Variation with number of atoms (4 nodes). (b) Variation with node count (512 atoms). The measurements in panels (a) and (b) are performed on a supercell of diamond-structure silicon.}
  \label{tab:load_balance}
  \begin{minipage}[t]{0.48\textwidth}
    \centering
    \begin{tabular}[t]{c c c}
    \hline
    \rule{0pt}{10ex}
    \begingroup
  \setbox0=\hbox{\shortstack{Number\\ of atoms}}%
  \setbox1=\hbox{\footnotesize(a)}%
  \makebox[\dimexpr \wd1 + 0.60em + \wd0\relax][l]{%
    \raisebox{\dimexpr (\ht0+\dp0-\ht1-\dp1)/2 + 3.00ex\relax}{\copy1}%
    \hspace{0.60em}%
    \copy0
  }%
\endgroup &
    \shortstack{Coefficient of\\variation for\\kernel time\\only (\%)} &
    \shortstack{Coefficient of\\variation including\\data transfer\\time (\%)} \\
    \hline
    216  & 0.27 & 0.61 \\
    384  & 0.39 & 0.58 \\
    512  & 0.17 & 0.22 \\
    864  & 0.54 & 0.51 \\
    1000 & 0.51 & 0.54 \\
    \hline
    \end{tabular}
  \end{minipage}
  \hfill
  \begin{minipage}[t]{0.48\textwidth}
    \centering
    \begin{tabular}[t]{c c c}
    \hline
    \rule{0pt}{10ex}
    \begingroup
  \setbox0=\hbox{\shortstack{Number\\ of nodes}}%
  \setbox1=\hbox{\footnotesize(b)}%
  \makebox[\dimexpr \wd1 + 0.60em + \wd0\relax][l]{%
    \raisebox{\dimexpr (\ht0+\dp0-\ht1-\dp1)/2 + 3.00ex\relax}{\copy1}%
    \hspace{0.60em}%
    \copy0
  }%
\endgroup &
    \shortstack{Coefficient of\\variation for\\kernel time\\only (\%)} &
    \shortstack{Coefficient of\\variation including\\data transfer\\time (\%)} \\
    \hline
    2 & 0.40 & 0.42 \\
    4 & 0.48 & 0.51 \\
    \hline
    \end{tabular}
  \end{minipage}
\end{table}

We analyze GPU utilization, defined based on the interval between the start of the first GPU event and the end of the last GPU event (the ``active time''). ``GPU utilization'' is defined as the kernel time divided by this active time. ``GPU utilization including data transfer time'' is defined as the total GPU processing time (kernel time plus transfer time) divided by the active time.

Table~\ref{tab:gpu_utility}(a) shows how utilization varies with the number of atoms (four nodes fixed). Utilization increases significantly as the number of atoms increases.
Table~\ref{tab:gpu_utility}(b) presents the utilization as the node count varies for a fixed system size (512 atoms). Utilization improves as fewer GPUs are used. The causes of these trends are discussed in Sect. 6.
\begin{table}[bt]
  \centering
  \caption{Change in GPU utilization for collinear DFT. (a) Variation with number of atoms (4 nodes). (b) Variation with node count (512 atoms). The measurements in panels (a) and (b) are performed on a supercell of diamond-structure silicon.}
  \label{tab:gpu_utility}
  \begin{minipage}[t]{0.495\textwidth}
    \centering
    \begin{tabular}[t]{c c c}
    \hline
    \noalign{\vskip 0.6ex}
    \begingroup
  \setbox0=\hbox{\shortstack{Number\\ of atoms}}%
  \setbox1=\hbox{\footnotesize(a)}%
  \makebox[\dimexpr \wd1 + 0.60em + \wd0\relax][l]{%
    \raisebox{\dimexpr (\ht0+\dp0-\ht1-\dp1)/2 + 0.10ex\relax}{\copy1}%
    \hspace{0.60em}%
    \copy0
  }%
\endgroup & GPU utilization (\%) & \shortstack{GPU utilization\\including\\data transfer\\time (\%)} \\
    \hline
    216 & 4.79 & 5.56 \\
    384 & 11.8 & 13.0 \\
    512 & 15.7 & 17.1 \\
    864 & 28.8 & 30.3 \\
    1000 & 32.1 & 33.5 \\
    \hline
    \end{tabular}
  \end{minipage}
  \hfill
  \begin{minipage}[t]{0.495\textwidth}
    \centering
    \begin{tabular}[t]{c c c}
    \hline
    \noalign{\vskip 0.6ex}
    \begingroup
  \setbox0=\hbox{\shortstack{Number\\ of nodes}}%
  \setbox1=\hbox{\footnotesize(b)}%
  \makebox[\dimexpr \wd1 + 0.60em + \wd0\relax][l]{%
    \raisebox{\dimexpr (\ht0+\dp0-\ht1-\dp1)/2 + 0.10ex\relax}{\copy1}%
    \hspace{0.60em}%
    \copy0
  }%
\endgroup & GPU utilization (\%) & \shortstack{GPU utilization\\including\\data transfer\\time (\%)} \\
    \hline
    2 & 23.7 & 24.7 \\
    4 & 15.7 & 17.1 \\
    \hline
    \end{tabular}
  \end{minipage}
\end{table}

\begin{figure}[bt]
  \centering
  \begin{minipage}[t]{0.48\textwidth}
    \centering
    \includegraphics[width=\linewidth]{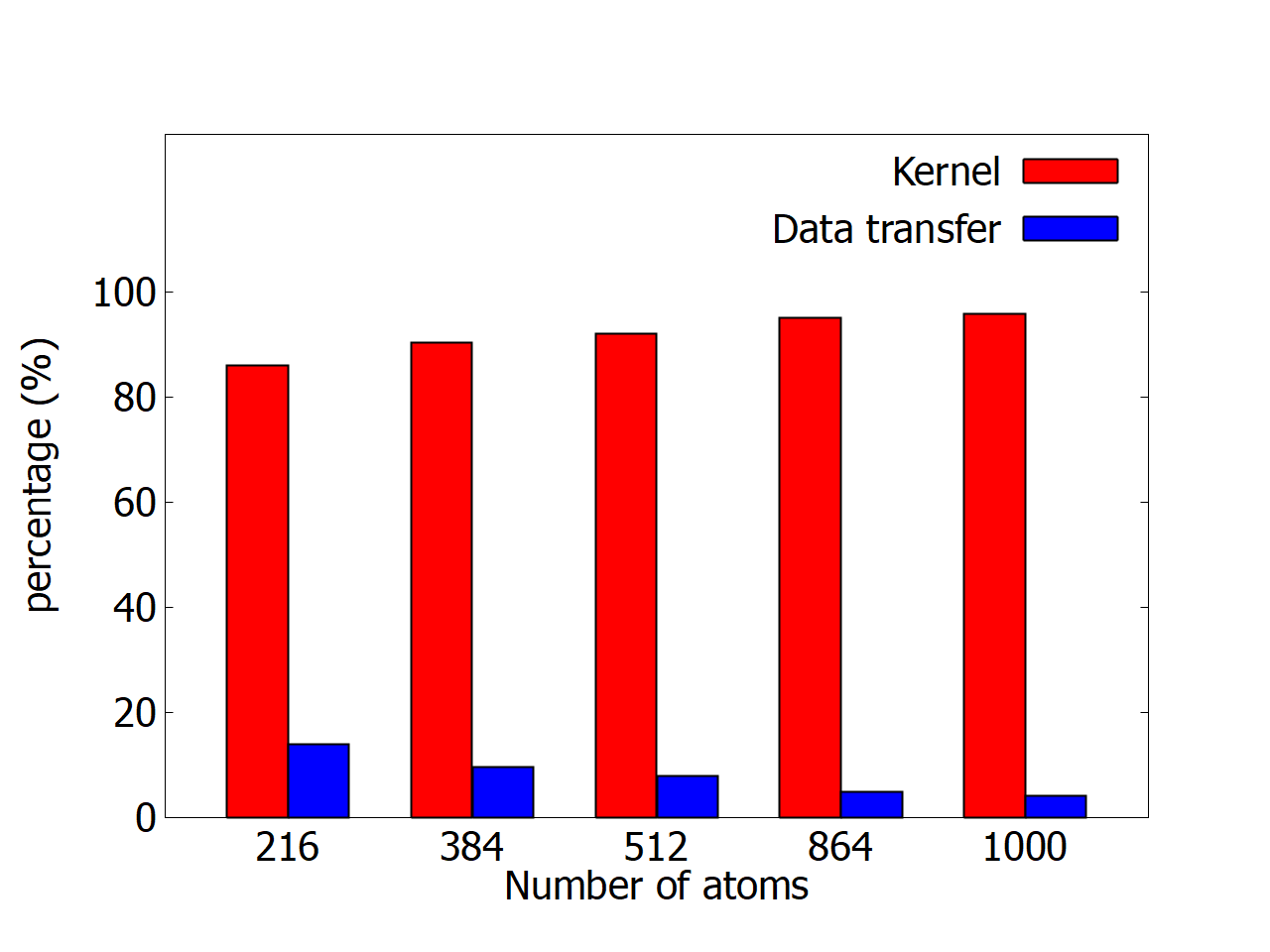}
    \footnotesize{(a)}
  \end{minipage}
  \hfill
  \begin{minipage}[t]{0.48\textwidth}
    \centering
    \includegraphics[width=\linewidth]{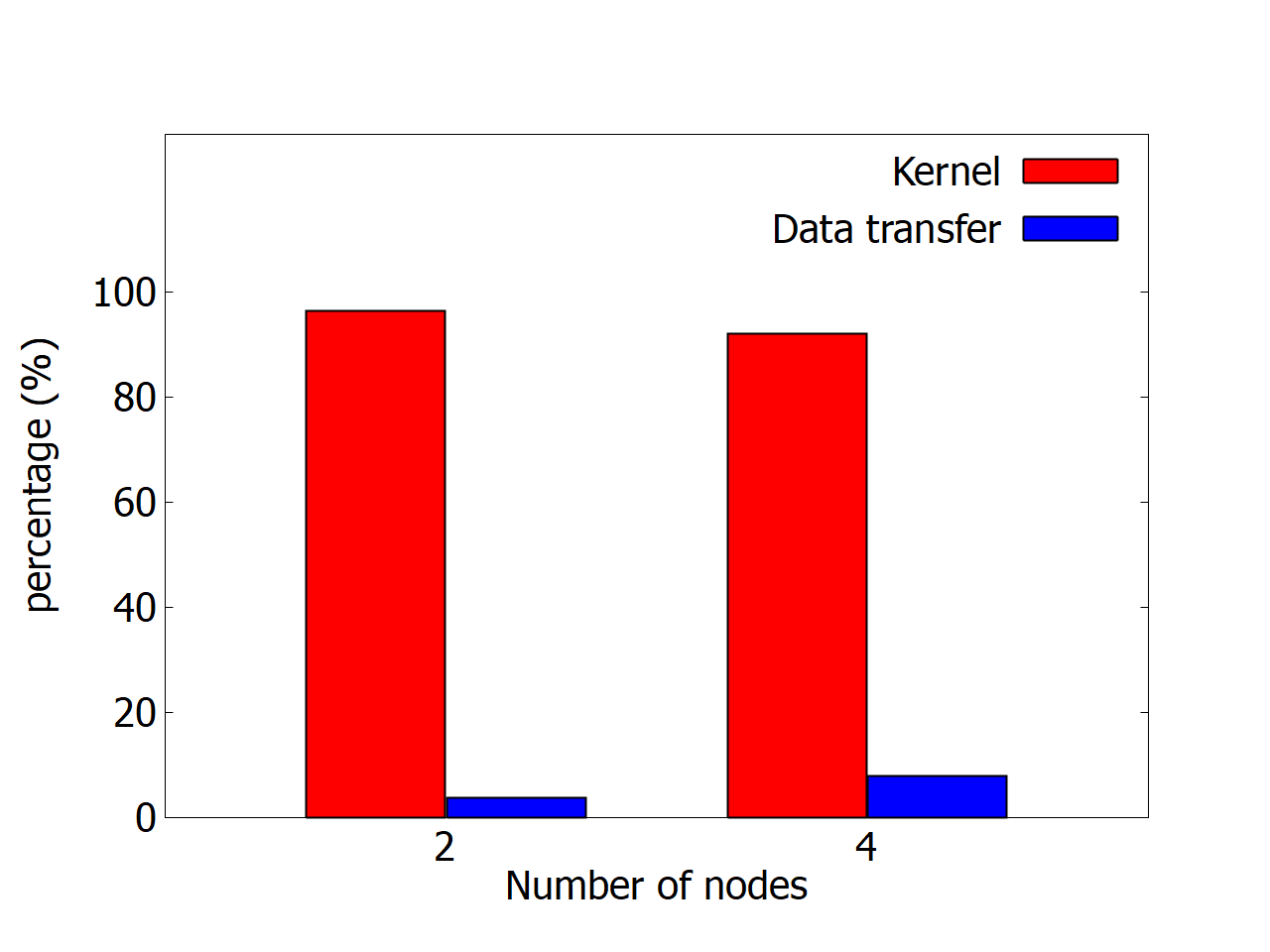}
    \footnotesize{(b)}
  \end{minipage}
  \caption{Fraction of host-device data transfer time in collinear DFT. (a) Variation with number of atoms (4 nodes). (b) Variation with node count (512 atoms). The measurements in panels (a) and (b) are performed on a supercell of diamond-structure silicon.}
  \label{fig:percentage}
\end{figure}

Figure~\ref{fig:percentage}(a) and Table~\ref{tab:percentage}(a) show the fraction of GPU time spent in kernels versus host-device transfers as a function of number of atoms (four nodes fixed). Kernel execution accounts for 86\% or more of the GPU processing time. The fraction of time spent on transfers decreases as the system size increases.
Figure~\ref{fig:percentage}(b) and Table~\ref{tab:percentage}(b) show these fractions as a function of node count (512 atoms fixed). Configurations with six or more nodes are excluded as only four GPUs are utilized. Kernel execution exceeds 90\% of the processing time. The trend shows that the data transfer fraction increases with the number of nodes. These trends regarding data transfer are discussed in Sect. 6.

\begin{table}[bt]
  \centering
  \caption{Fraction of host-device data transfer time in collinear DFT. (a) Variation with number of atoms. (b) Variation with node count. The measurements in panels (a) and (b) are performed on the same system as in Fig.~\ref{fig:percentage}.}
  \label{tab:percentage}
  \begin{minipage}[t]{0.48\textwidth}
    \centering
    \begin{tabular}[t]{c c c c}
    \hline
    \rule{0pt}{7ex}
    \begingroup
  \setbox0=\hbox{\shortstack{Number\\ of atoms}}%
  \setbox1=\hbox{\footnotesize(a)}%
  \makebox[\dimexpr \wd1 + 0.60em + \wd0\relax][l]{%
    \raisebox{\dimexpr (\ht0+\dp0-\ht1-\dp1)/2 + 1.20ex\relax}{\copy1}%
    \hspace{0.60em}%
    \copy0
  }%
\endgroup &
    \shortstack{Fraction of\\kernel time (\%)} &
    \shortstack{Fraction of\\data transfer\\time (\%)} \\
    \hline
    216 &  86.14 & 13.86 \\
    384 &  90.37 & 9.63 \\
    512 &  92.10 & 7.90 \\
    864 &  95.15 & 4.85 \\
    1000 & 95.80 & 4.20 \\
    \hline
    \end{tabular}
  \end{minipage}
  \hfill
  \begin{minipage}[t]{0.48\textwidth}
    \centering
    \begin{tabular}[t]{c c c}
    \hline
    \rule{0pt}{7ex}
    \begingroup
  \setbox0=\hbox{\shortstack{Number\\ of nodes}}%
  \setbox1=\hbox{\footnotesize(b)}%
  \makebox[\dimexpr \wd1 + 0.60em + \wd0\relax][l]{%
    \raisebox{\dimexpr (\ht0+\dp0-\ht1-\dp1)/2 + 1.20ex\relax}{\copy1}%
    \hspace{0.60em}%
    \copy0
  }%
\endgroup &
    \shortstack{Fraction of\\kernel time (\%)} &
    \shortstack{Fraction of\\data transfer\\time (\%)} \\
    \hline
    2 & 96.35 &  3.65 \\
    4 & 92.10 &  7.90 \\
    \hline
    \end{tabular}
  \end{minipage}
\end{table}

Table~\ref{tab:percentage_data}(a) details the host-device data transfer volume, transfer time, and total GPU processing time (kernel time plus transfer time) versus the number of atoms (four nodes fixed). Both transfer volume and time increase with the number of atoms.
Table~\ref{tab:percentage_data}(b) shows these metrics versus the number of nodes (512 atoms fixed). When reducing from four nodes (4 GPUs) to two nodes (2 GPUs), each GPU handles two MPI processes (two $\mathbf{k}$-points). Consequently, the transfer volume per GPU doubles, and the transfer time nearly doubles. The total GPU processing time increases by nearly a factor of four.
A detailed discussion of this point is provided in Sect. 6.

\begin{table}[bt]
  \centering
  \caption{Host-device data transfer volume, transfer time, and GPU processing time in collinear DFT. (a) Variation with number of atoms. (b) Variation with node count. The measurements in panels (a) and (b) are performed on the same system as in Fig.~\ref{fig:percentage}.}
  \label{tab:percentage_data}
  \setlength{\tabcolsep}{4pt}
  \begin{minipage}[t]{0.495\textwidth}
    \centering
    \begin{tabular}[t]{c c c c}
    \hline
    \noalign{\vskip 0.4ex}
    \rule{0pt}{5ex}
    \begingroup
  \setbox0=\hbox{\shortstack{Number\\ of atoms}}%
  \setbox1=\hbox{\footnotesize(a)}%
  \makebox[\dimexpr \wd1 + 0.60em + \wd0\relax][l]{%
    \raisebox{\dimexpr (\ht0+\dp0-\ht1-\dp1)/2 + 1.20ex\relax}{\copy1}%
    \hspace{0.60em}%
    \copy0
  }%
\endgroup &
    \shortstack{Data\\transfer\\volume (MB)} &
    \shortstack{Data\\transfer\\time (sec)} &
    \shortstack{GPU\\processing\\time (sec)} \\
    \hline
    216 &  15896 & 0.526 & 3.79 \\
    384 &  50240 & 1.50 & 15.6 \\
    512 &  89315 & 2.36 & 29.9 \\
    864 &  254337 & 5.95 & 122.7 \\
    1000 & 340707 & 7.66 & 182.3 \\
    \hline
    \end{tabular}
  \end{minipage}
  \hfill
  \begin{minipage}[t]{0.495\textwidth}
    \centering
    \begin{tabular}[t]{c c c c}
    \hline
    \noalign{\vskip 0.4ex}
    \rule{0pt}{5ex}
    \begingroup
  \setbox0=\hbox{\shortstack{Number\\ of nodes}}%
  \setbox1=\hbox{\footnotesize(b)}%
  \makebox[\dimexpr \wd1 + 0.60em + \wd0\relax][l]{%
    \raisebox{\dimexpr (\ht0+\dp0-\ht1-\dp1)/2 + 1.20ex\relax}{\copy1}%
    \hspace{0.60em}%
    \copy0
  }%
\endgroup &
    \shortstack{Data\\transfer\\volume (MB)} &
    \shortstack{Data\\transfer\\time (sec)} &
    \shortstack{GPU\\processing\\time (sec)} \\
    \hline
    2 & 178630 & 4.42 & 120.8 \\
    4 & 89315 & 2.31 & 29.89 \\
    \hline
    \end{tabular}
  \end{minipage}
\end{table}

\subsection{Noncollinear DFT results}
We now examine the results for noncollinear DFT. The baseline CPU-only performance is measured for a 640-atom silicon system (8,320 basis functions) on eight nodes (384 CPU cores). Table~\ref{tab:ncdft_cpu_time} summarizes the wall time breakdown.
\begin{table}[bt]
  \centering
  \caption{Breakdown of noncollinear DFT wall time for the CPU-only calculation. Measurements are performed on a 640-atom supercell of diamond-structure silicon.}
  \label{tab:ncdft_cpu_time}
  \begin{tabular}{c c c}
  \hline
  Routine & Time (sec) & Proportion (\%) \\
  \hline
  Set OLP Kin         & 0.694  & 0.03 \\
  Set Nonlocal        & 2.236  & 0.11 \\
  Set ProExpn VNA     & 6.737  & 0.33 \\
  Set Hamiltonian     & 56.32  & 2.78 \\
  Diagonalization     & 1936   & 95.4 \\
  Mixing DM           & 0.984  & 0.05 \\
  Force               & 4.695  & 0.23 \\
  Total Energy        & 2.156  & 0.11 \\
  Set Density Grid    & 8.909  & 0.44 \\
  FFT(2D) Density     & 0.405  & 0.02 \\
  Others              & 10.343 & 0.51 \\
  Total               & 2029   & 100.0 \\
  \hline
  \end{tabular}
\end{table}
Similar to the collinear case, diagonalization dominates the execution, accounting for 95.4\% of the total wall time.
Table~\ref{tab:breakdown_ncdft} provides a detailed breakdown of the diagonalization time (maximum across 384 MPI processes) for the first SCF iteration (a) and subsequent iterations (b). In the first iteration, eigenvalue solutions (Parts 3-1 and 3-5) account for 60.9\% of the time, and matrix multiplications (Parts 3-3 and 3-6) take 25.5\%. In subsequent iterations, where Parts 3-1 and 3-2 are skipped, the eigenvalue solution (Part 3-5) takes 56.4\%, and matrix multiplications take 32.6\%.
\begin{table}[bt]
    \centering
    \caption{Breakdown of the noncollinear diagonalization time (CPU-only). (a) First SCF iteration. (b) SCF iterations 2--22 (average). Measurements are performed on a 640-atom supercell of diamond-structure silicon.}
    \label{tab:breakdown_ncdft}
    \begin{minipage}[t]{0.48\textwidth}
      \centering
      \begin{tabular}[t]{c c c}
        \toprule
        \begingroup
  \setbox0=\hbox{\shortstack{Routine}}%
  \setbox1=\hbox{\footnotesize(a)}%
  \makebox[\dimexpr \wd1 + 0.60em + \wd0\relax][l]{%
    \raisebox{\dimexpr (\ht0+\dp0-\ht1-\dp1)/2 + 0.10ex\relax}{\copy1}%
    \hspace{0.60em}%
    \copy0
  }%
\endgroup & Time (sec) & Proportion (\%) \\
        \midrule
        Part 1     & 0.609  & 0.56 \\
        Part 2     & 6.307  & 5.83 \\
        Part 3-1   & 19.77  & 18.3 \\
        Part 3-2   & 1.685  & 1.56 \\
        Part 3-3   & 12.95  & 12.0 \\
        Part 3-4   & 1.956  & 1.81 \\
        Part 3-5   & 46.12  & 42.6 \\
        Part 3-6   & 14.64  & 13.5 \\
        Part 4     & 0.0194 & 0.02 \\
        Part 5     & 4.171  & 3.85 \\
        Total      & 108.2  & 100.0 \\
        \bottomrule
      \end{tabular}
    \end{minipage}
    \hfill
    \begin{minipage}[t]{0.48\textwidth}
      \centering
      \begin{tabular}[t]{c c c}
        \toprule
        \begingroup
  \setbox0=\hbox{\shortstack{Routine}}%
  \setbox1=\hbox{\footnotesize(b)}%
  \makebox[\dimexpr \wd1 + 0.60em + \wd0\relax][l]{%
    \raisebox{\dimexpr (\ht0+\dp0-\ht1-\dp1)/2 + 0.10ex\relax}{\copy1}%
    \hspace{0.60em}%
    \copy0
  }%
\endgroup & Time (sec) & Proportion (\%) \\
        \midrule
        Part 1     & 0.000  & 0.00 \\
        Part 2     & 3.170  & 3.76 \\
        Part 3-1   & 0.000  & 0.00 \\
        Part 3-2   & 0.000  & 0.00 \\
        Part 3-3   & 12.96  & 15.4 \\
        Part 3-4   & 1.863  & 2.21 \\
        Part 3-5   & 47.55  & 56.4 \\
        Part 3-6   & 14.50  & 17.2 \\
        Part 4     & 0.0371 & 0.04 \\
        Part 5     & 4.262  & 5.05 \\
        Total      & 84.33  & 100.0 \\
        \bottomrule
      \end{tabular}
    \end{minipage}
\end{table}

When eight GPUs are added to the configuration, the wall-time breakdown changes as shown in Table~\ref{tab:ncdft_cpu_time_gpu}. Diagonalization remains the dominant phase, accounting for 86.8\% of the total time.
\begin{table}[bt]
  \centering
  \caption{Breakdown of noncollinear DFT wall time for the GPU-accelerated calculation. Measurements are performed on a 640-atom supercell of diamond-structure silicon.}
  \label{tab:ncdft_cpu_time_gpu}
  \begin{tabular}{c c c}
  \hline
  Routine & Time (sec) & Proportion (\%) \\
  \hline
  Set OLP Kin        & 0.693  & 0.09 \\
  Set Nonlocal       & 2.208  & 0.30 \\
  Set ProExpn VNA    & 6.771  & 0.92 \\
  Set Hamiltonian    & 60.27  & 8.18 \\
  Diagonalization    & 639.9  & 86.8 \\
  Mixing DM          & 1.123  & 0.15 \\
  Force              & 4.885  & 0.66 \\
  Total Energy       & 2.104  & 0.29 \\
  Set Density Grid   & 8.960  & 1.22 \\
  FFT(2D) Density    & 0.611  & 0.08 \\
  Others             & 9.645  & 1.31 \\
  Total              & 737.2  & 100.0 \\
  \hline
  \end{tabular}
\end{table}

Table~\ref{tab:breakdown_ncdft_gpu}(a) details the diagonalization breakdown during the first SCF iteration with GPU acceleration. The combined share of eigenvalue solutions (36.2\%) and matrix multiplications (23.9\%) is 60.1\%, a significant reduction from 86.4\% in the CPU-only calculation. However, similar to the collinear results, the time spent in earlier stages increases due to forced serialization. Part 2 (building the Hamiltonian blocks $\mathbf{H}^{\alpha\alpha}, \mathbf{H}^{\alpha\beta}, \mathbf{H}^{\beta\alpha}, \mathbf{H}^{\beta\beta}$) increases because it is not MPI-parallelized in the GPU implementation. 
Furthermore, Part 3-6 includes the overhead for redistributing matrices across MPI ranks, accounting for a non-negligible 8.63\%.

The achieved speedups for the eigenvalue solutions are 5.44 times (Part 3-1) and 4.40 times (Part 3-5). Matrix multiplication (Part 3-3) achieves a 15.6 times speedup. (The high speedup for Part 3-3 compared to the collinear Part 2-3 is partly because host-device transfer time is excluded from this specific measurement).
In subsequent SCF iterations [Table~\ref{tab:breakdown_ncdft_gpu}(b)], Part 3-5 takes 41.0\%, and Parts 3-3 and 3-6 take 21.8\% of the time. The combined share of these steps is 62.8\%, reduced from 89.0\% (CPU-only). The speedups are 14.6 times (Part 3-3) and 4.30 times (Part 3-5). As observed in the previous case, the overhead for this redistribution remains substantial, now reaching 10.0\%.

\begin{table}[bt]
    \centering
    \caption{Breakdown of the noncollinear diagonalization time for the GPU-accelerated calculation. (a) First SCF iteration. (b) SCF iterations 2--22 (average). Measurements are performed on a 640-atom supercell of diamond-structure silicon.}
    \label{tab:breakdown_ncdft_gpu}
    \begin{minipage}[t]{0.48\textwidth}
      \centering
      \begin{tabular}[t]{c c c}
        \toprule
        \begingroup
  \setbox0=\hbox{\shortstack{Routine}}%
  \setbox1=\hbox{\footnotesize(a)}%
  \makebox[\dimexpr \wd1 + 0.60em + \wd0\relax][l]{%
    \raisebox{\dimexpr (\ht0+\dp0-\ht1-\dp1)/2 + 0.10ex\relax}{\copy1}%
    \hspace{0.60em}%
    \copy0
  }%
\endgroup & Time (sec) & Proportion (\%) \\
        \midrule
        Part 1   & 0.9258 & 2.37 \\
        Part 2   & 8.234  & 21.1 \\
        Part 3-1 & 3.636  & 9.31 \\
        Part 3-2 & 2.218  & 5.68 \\
        Part 3-3 & 0.8314 & 2.13 \\
        Part 3-4 & 0.0069 & 0.02 \\
        Part 3-5 & 10.49  & 26.9 \\
        Part 3-6 & 8.513  & 21.8 \\
        (MPI)    & (3.369)   & (8.63) \\
        Part 4   & 0.021  & 0.05 \\
        Part 5   & 4.178  & 10.7 \\
        Total    & 39.06  & 100.0 \\
        \bottomrule
      \end{tabular}
    \end{minipage}
    \hfill
    \begin{minipage}[t]{0.48\textwidth}
      \centering
      \begin{tabular}[t]{c c c}
        \toprule
        \begingroup
  \setbox0=\hbox{\shortstack{Routine}}%
  \setbox1=\hbox{\footnotesize(b)}%
  \makebox[\dimexpr \wd1 + 0.60em + \wd0\relax][l]{%
    \raisebox{\dimexpr (\ht0+\dp0-\ht1-\dp1)/2 + 0.10ex\relax}{\copy1}%
    \hspace{0.60em}%
    \copy0
  }%
\endgroup & Time (sec) & Proportion (\%) \\
        \midrule
        Part 1   & 0.0000 & 0.00 \\
        Part 2   & 5.798  & 21.5 \\
        Part 3-1 & 0.0000 & 0.00 \\
        Part 3-2 & 0.0000 & 0.00 \\
        Part 3-3 & 0.8885 & 3.29 \\
        Part 3-4 & 0.0067 & 0.02 \\
        Part 3-5 & 11.07  & 41.0 \\
        Part 3-6 & 4.988  & 18.5 \\
        (MPI)    & (2.708)  & (10.0) \\
        Part 4   & 0.020  & 0.07 \\
        Part 5   & 4.219  & 15.6 \\
        Total    & 26.99  & 100.00 \\
        \bottomrule
      \end{tabular}
    \end{minipage}
\end{table}

\begin{figure}[bt]
  \centering
  \includegraphics[width=0.49\textwidth]{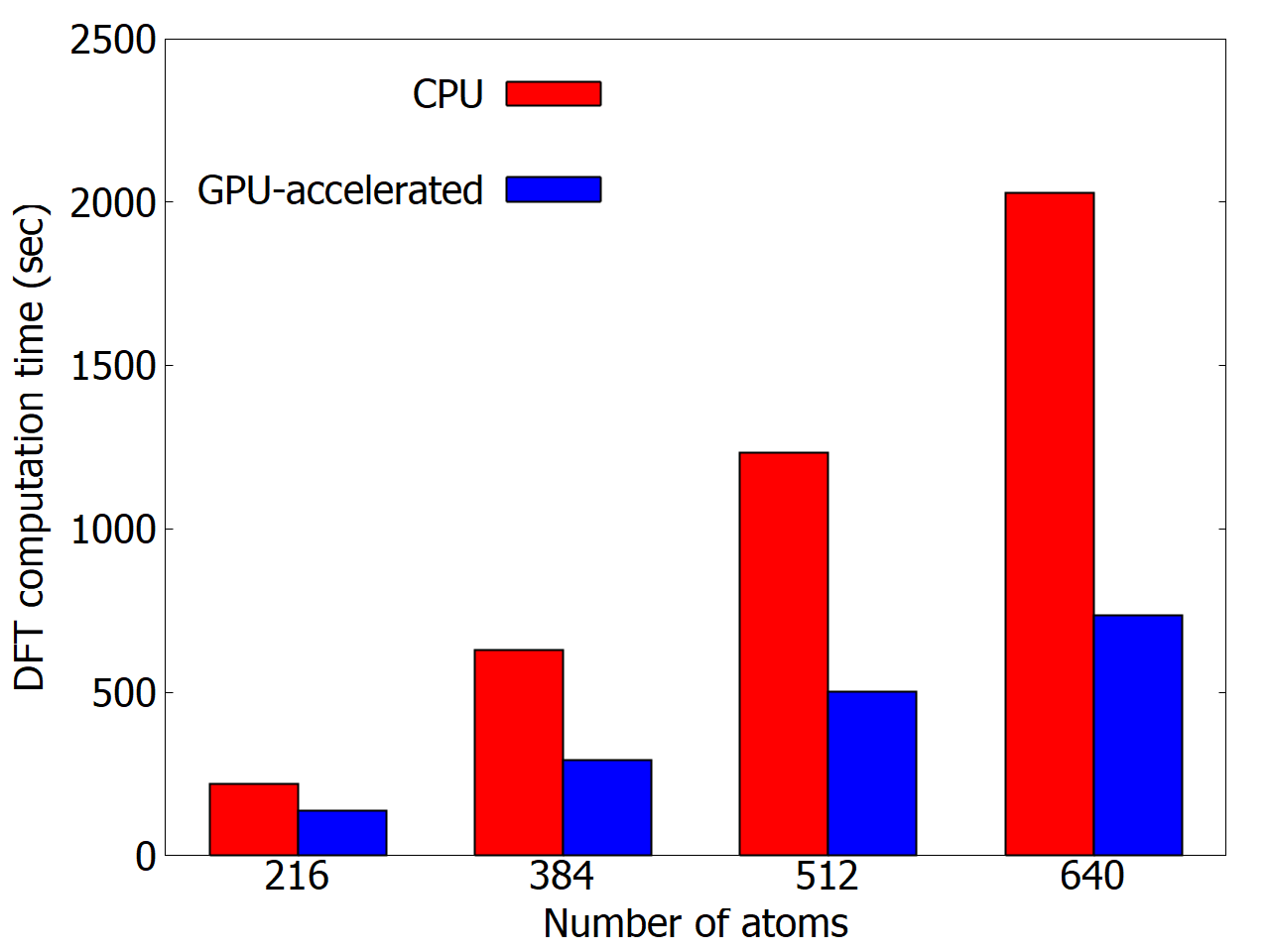}
  \caption{Relationship between number of atoms and total noncollinear DFT wall time (Eight nodes).}
  \label{fig:atoms_noncol_bar_graph}
\end{figure}

\begin{figure}[bt]
  \centering
  \includegraphics[width=0.49\textwidth]{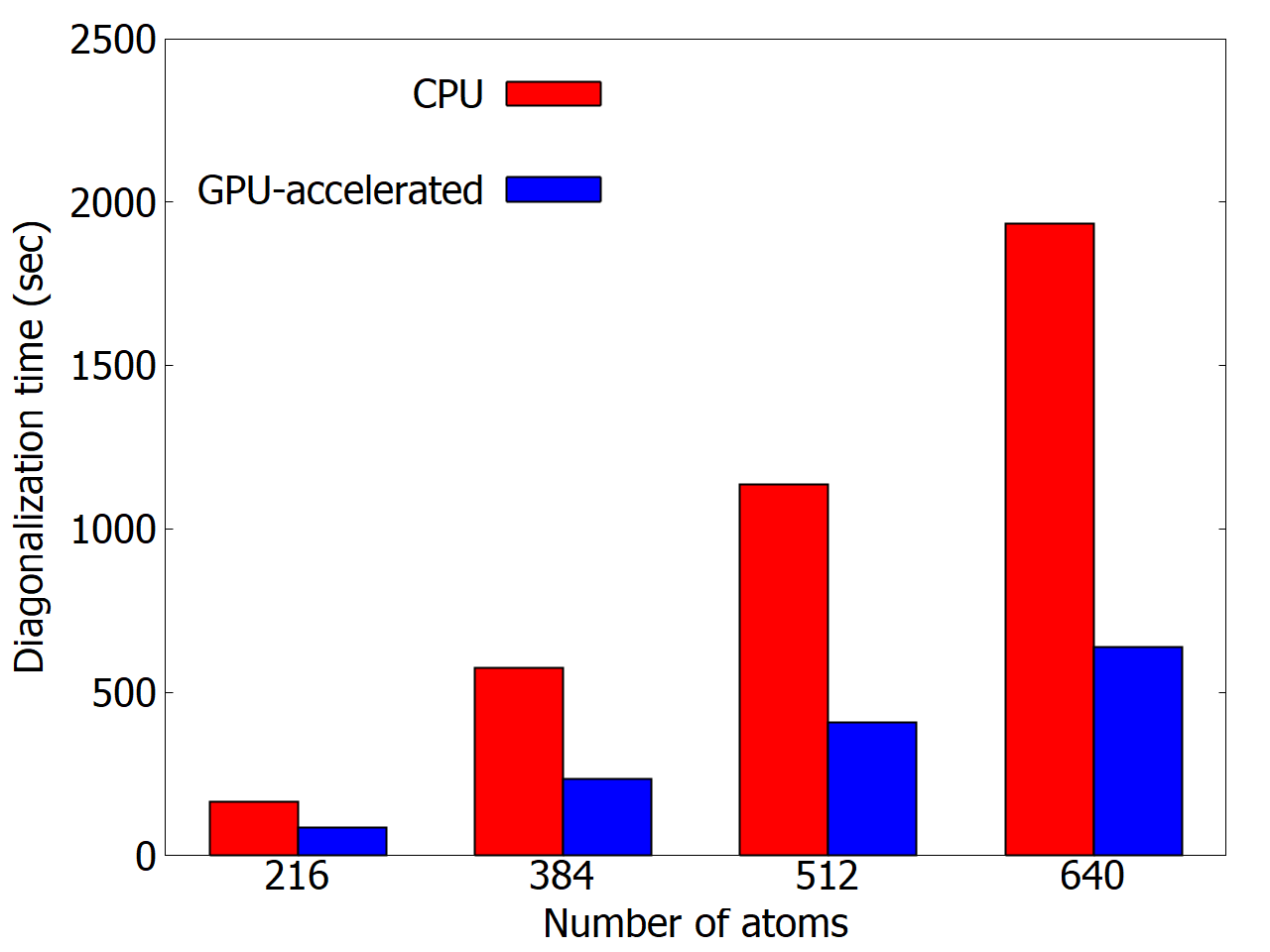}
  \caption{Relationship between number of atoms and diagonalization time in noncollinear DFT (Eight nodes).}
  \label{fig:atoms_noncol_diag_bar_graph}
\end{figure}

Figure~\ref{fig:atoms_noncol_bar_graph} and Table~\ref{tab:atoms_noncol} compare the total noncollinear DFT wall times versus number of atoms on eight nodes. The GPU implementation consistently outperforms the CPU-only calculations across all system sizes.
\begin{table}[bt]
  \centering
  \caption{Variation of total noncollinear DFT wall time (s) with number of atoms (Eight nodes).}
  \label{tab:atoms_noncol}
  \begin{tabular}{c c c c}
  \hline
  \rule{0pt}{5ex}
  \shortstack{Number\\of atoms} & CPU & \shortstack{GPU-\\accelerated} & \shortstack{Speed-up\\(times)} \\
  \hline
   216 &  220.3 & 139.2 & 1.58 \\
   384 &  629.3 & 291.7 & 2.16 \\
   512 & 1232.6 & 502.0 & 2.46 \\
   640 & 2029.3 & 737.2 & 2.75 \\
  \hline
  \end{tabular}
\end{table}
For the largest system tested (640 atoms), a speedup of 2.75 times is achieved.
Figure~\ref{fig:atoms_noncol_diag_bar_graph} and Table~\ref{tab:atoms_noncol_diag} show the corresponding diagonalization times.
\begin{table}[bt]
  \centering
  \caption{Variation of diagonalization time (s) with number of atoms for noncollinear DFT (Eight nodes).}
  \label{tab:atoms_noncol_diag}
  \begin{tabular}{c c c c}
  \hline
  \rule{0pt}{5ex}
  \shortstack{Number\\of atoms} & CPU & \shortstack{GPU-\\accelerated} & \shortstack{Speed-up\\(times)} \\
  \hline
   216 &  165.5 &  86.0 & 1.92 \\
   384 &  573.8 & 236.2 & 2.43 \\
   512 & 1137.4 & 409.3 & 2.78 \\
   640 & 1935.8 & 639.9 & 3.03 \\
  \hline
  \end{tabular}
\end{table}
The speedup for the diagonalization phase reaches 3.03 times for the 640-atom system. As observed in the collinear case, this speedup relative to the GPU's theoretical potential is discussed in Sect. 6.

\begin{figure}[bt]
  \centering
  \begin{minipage}[t]{0.48\textwidth}
    \centering
    \includegraphics[width=\linewidth]{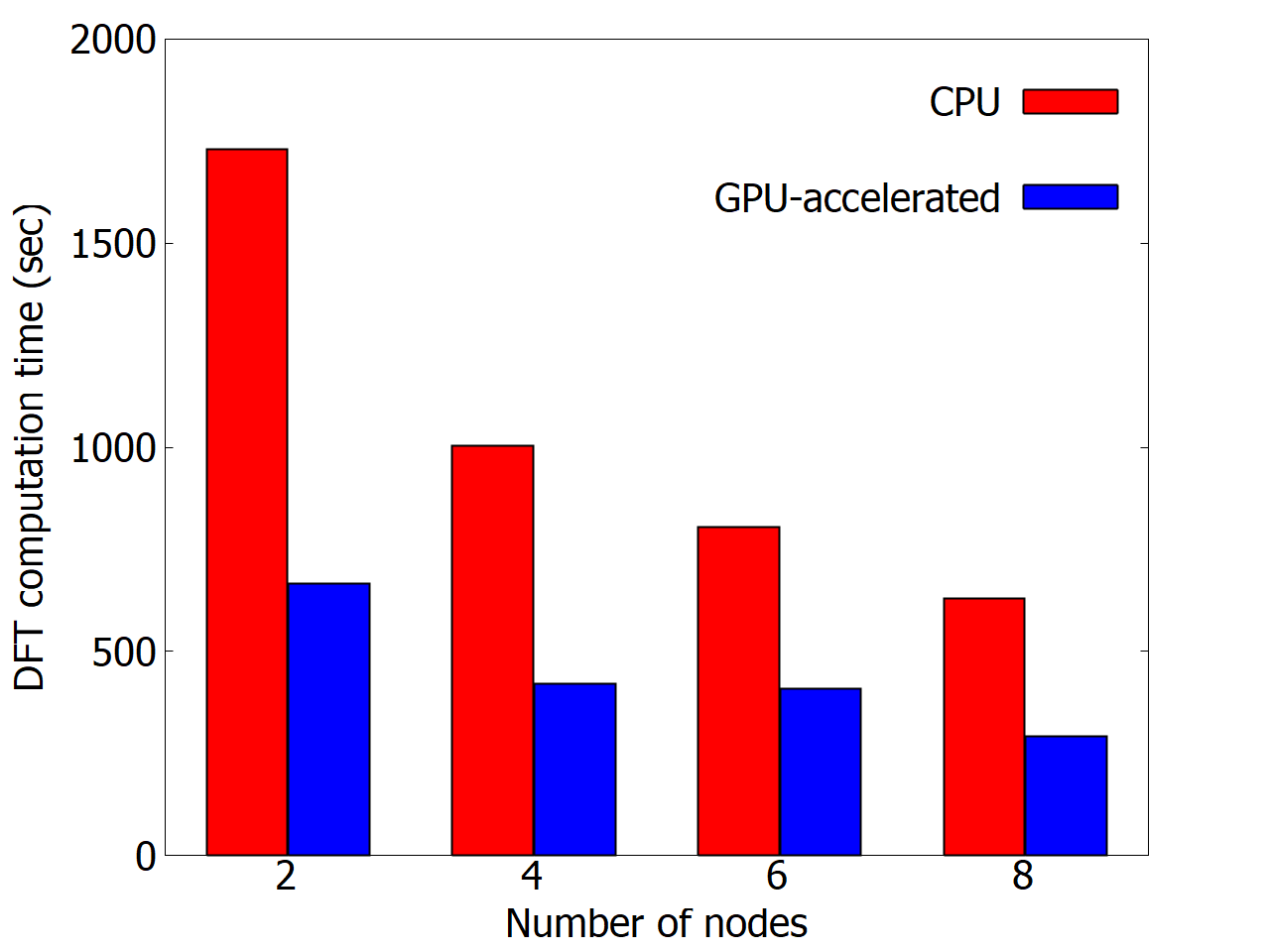}
    \footnotesize{(a)}
  \end{minipage}
  \hfill
  \begin{minipage}[t]{0.48\textwidth}
    \centering
    \includegraphics[width=\linewidth]{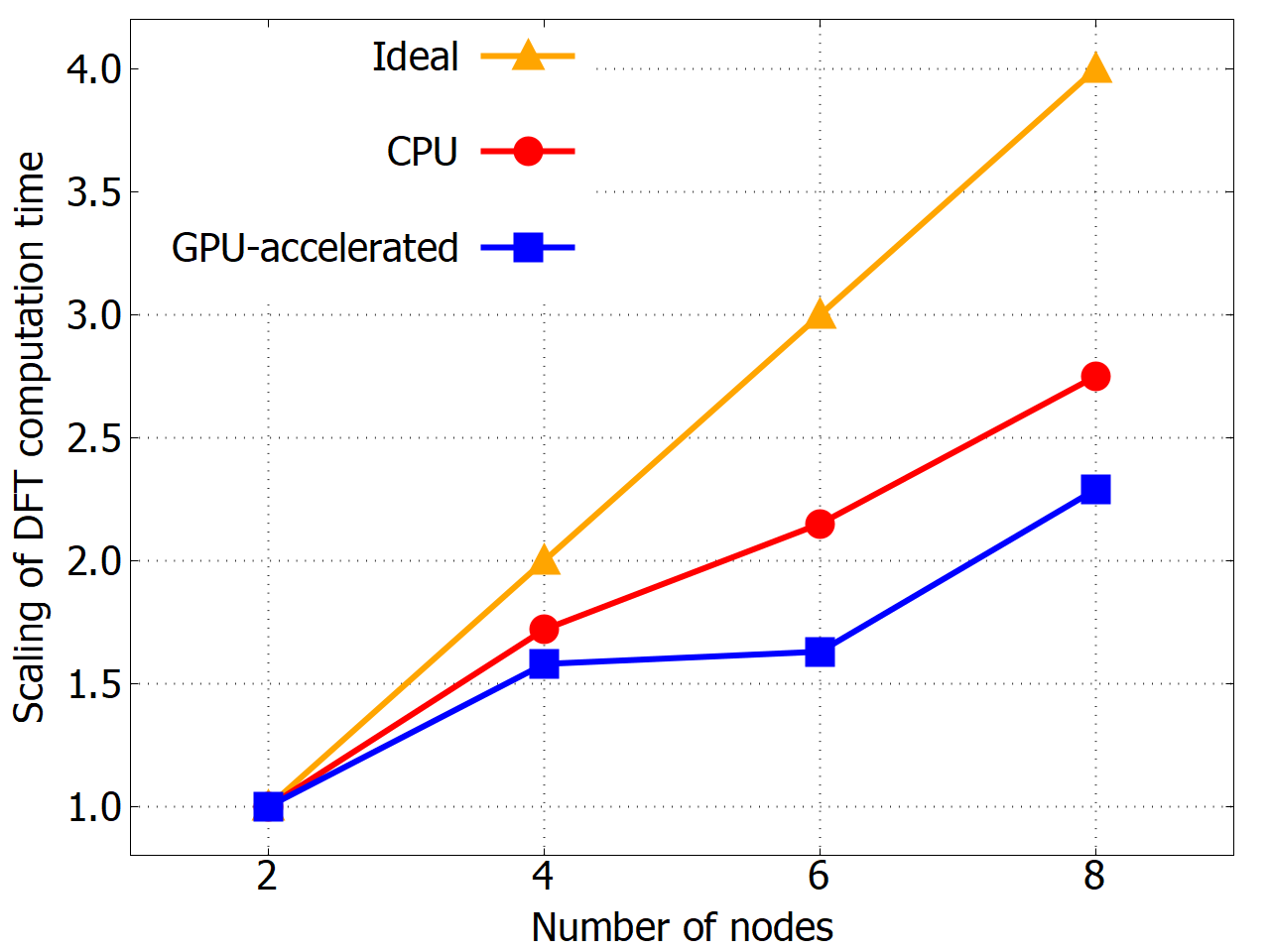}
    \footnotesize{(b)}
  \end{minipage}
  \caption{Comparison of performance for noncollinear DFT as the node count increases. (a) Comparison of wall time. (b) Comparison of scaling (relative to 2 nodes). Measurements are performed on a 384-atom supercell of diamond-structure silicon.}
  \label{fig:nodes_noncol}
\end{figure}

We next examine the scaling behavior for a fixed number of atoms (384 atoms) while varying the node count from 2 to 8. Figure~\ref{fig:nodes_noncol} compares the DFT calculation times (a) and the scaling behavior relative to two nodes (b). Table~\ref{tab:nodes_noncol} provides the numerical data.
\begin{table}[bt]
  \centering
  \caption{Variation of total DFT wall time (s) and scaling with node count for noncollinear DFT. (a) Wall time. (b) Scaling. The system is the same as for Fig.~\ref{fig:nodes_noncol}.}
  \label{tab:nodes_noncol}
  \begin{minipage}[t]{0.48\textwidth}
    \centering
    \begin{tabular}[t]{c c c c}
    \hline
    \rule{0pt}{5ex}
    \begingroup
  \setbox0=\hbox{\shortstack{Number\\ of nodes}}%
  \setbox1=\hbox{\footnotesize(a)}%
  \makebox[\dimexpr \wd1 + 0.60em + \wd0\relax][l]{%
    \raisebox{\dimexpr (\ht0+\dp0-\ht1-\dp1)/2 + 0.10ex\relax}{\copy1}%
    \hspace{0.60em}%
    \copy0
  }%
\endgroup & CPU & \shortstack{GPU-\\accelerated} & \shortstack{Speed-up\\(times)} \\
    \hline
    2 & 1730.8 & 666.9 & 2.60 \\
    4 & 1003.7 & 421.4 & 2.38 \\
    6 & 804.7  & 410.1 & 1.96 \\
    8 & 629.3  & 291.7 & 2.16 \\
    \hline
    \end{tabular}
  \end{minipage}
  \hfill
  \begin{minipage}[t]{0.48\textwidth}
    \centering
    \begin{tabular}[t]{c c c}
    \hline
    \begingroup
  \setbox0=\hbox{\shortstack{Number of nodes}}%
  \setbox1=\hbox{\footnotesize(b)}%
  \makebox[\dimexpr \wd1 + 0.60em + \wd0\relax][l]{%
    \raisebox{\dimexpr (\ht0+\dp0-\ht1-\dp1)/2 + 0.10ex\relax}{\copy1}%
    \hspace{0.60em}%
    \copy0
  }%
\endgroup & CPU scaling & GPU scaling \\
    \hline
    2 & 1.00 & 1.00 \\
    4 & 1.72 & 1.58 \\
    6 & 2.15 & 1.63 \\
    8 & 2.75 & 2.29 \\
    \hline
    \noalign{\vskip 0.6ex}
    \end{tabular}
  \end{minipage}
\end{table}
The trends observed are similar to the collinear results. The GPU acceleration factor decreases as resources increase: a 2.60 times speedup is achieved on 2 nodes, dropping to 2.16 times on 8 nodes. The CPU-only configuration again demonstrates superior scaling (2.75 times speedup from 2 to 8 nodes) compared to the GPU-accelerated case (2.29 times speedup).

\begin{figure}[bt]
  \centering
  \begin{minipage}[t]{0.48\textwidth}
    \centering
    \includegraphics[width=\linewidth]{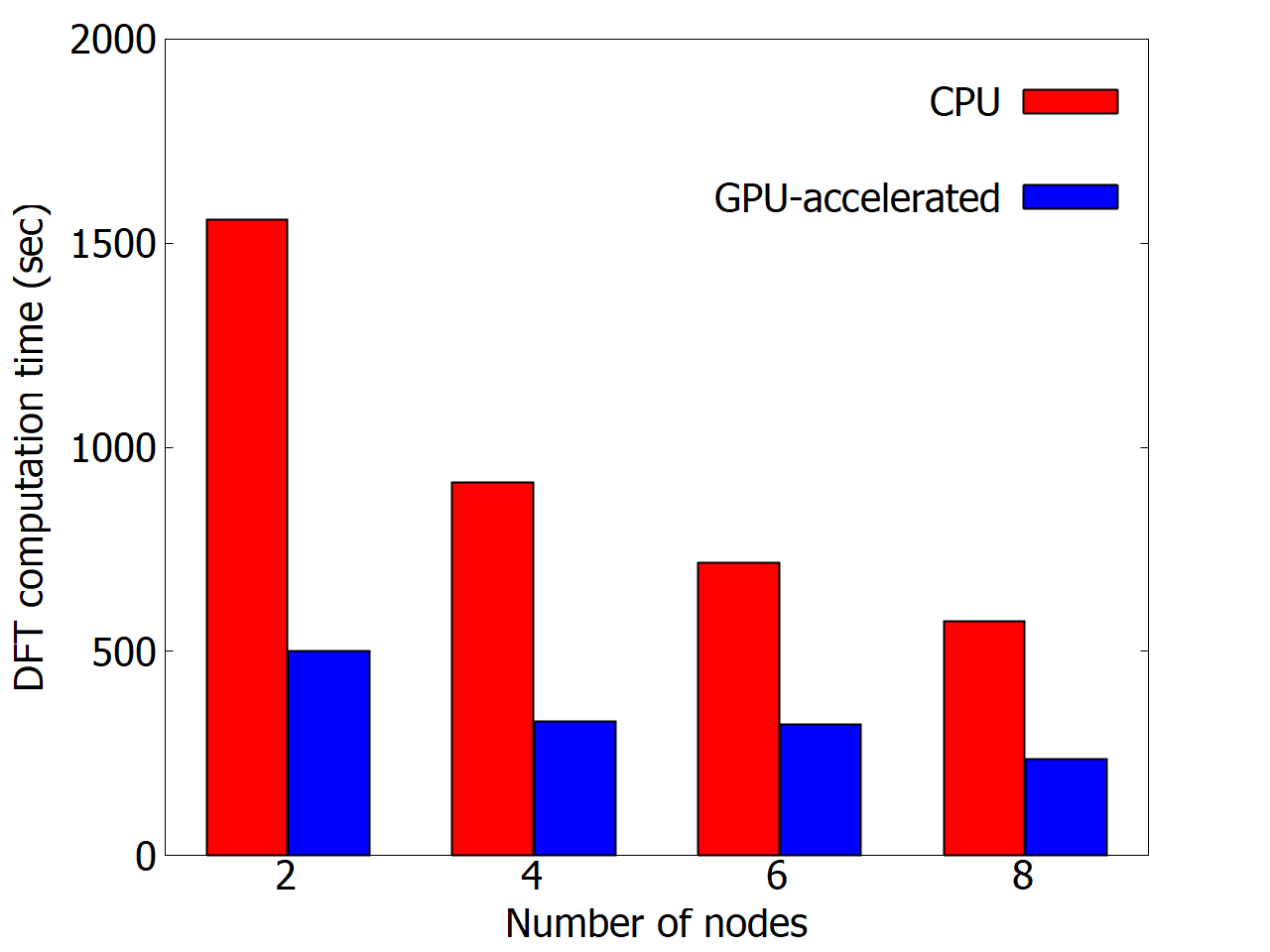}
    \footnotesize{(a)}
  \end{minipage}
  \hfill
  \begin{minipage}[t]{0.48\textwidth}
    \centering
    \includegraphics[width=\linewidth]{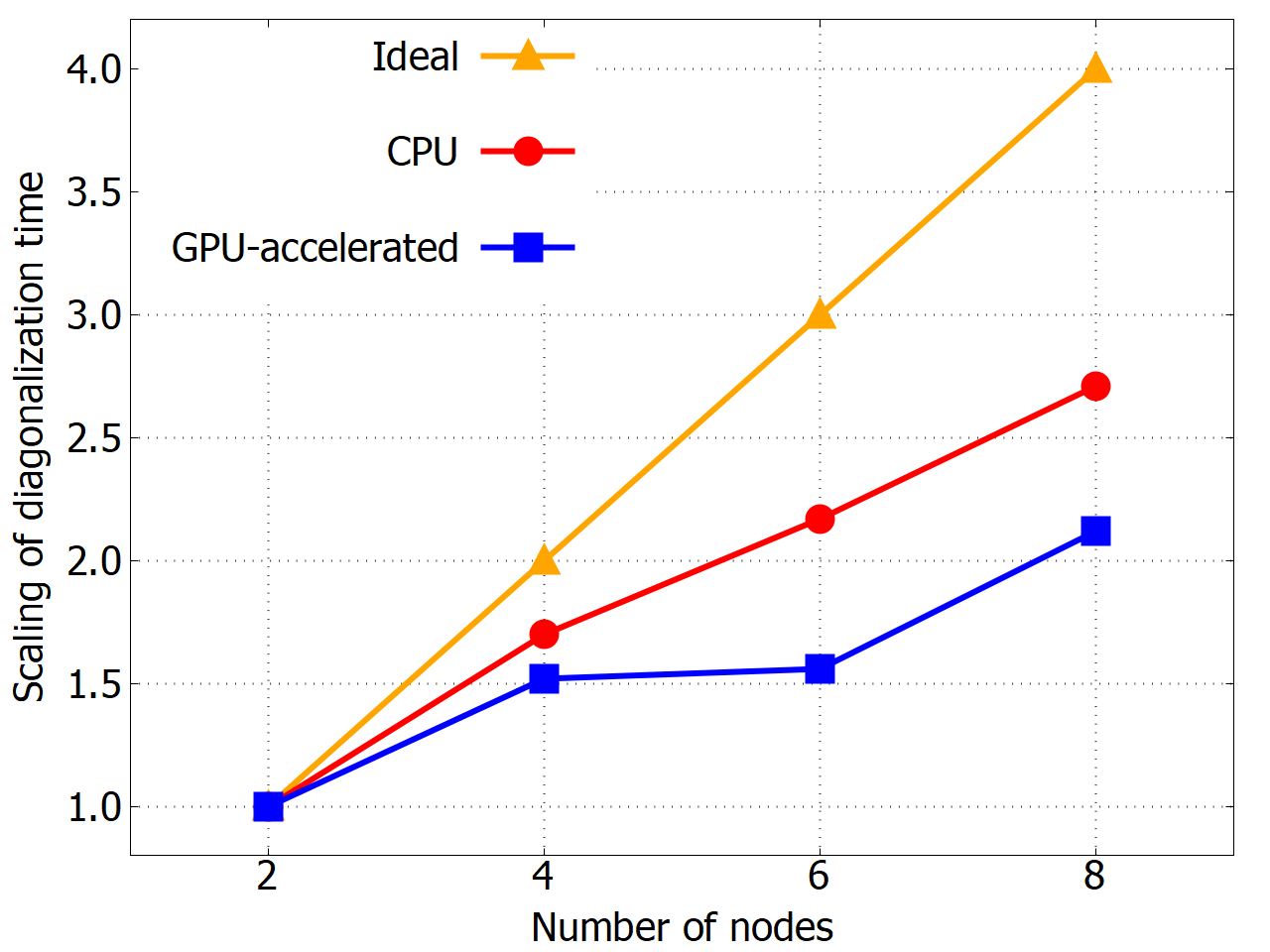}
    \footnotesize{(b)}
  \end{minipage}
  \caption{Comparison of diagonalization performance for noncollinear DFT as the node count increases. (a) Diagonalization time. (b) Scaling comparison. The system is the same as for Fig.~\ref{fig:nodes_noncol}.}
  \label{fig:nodes_noncol_diag}
\end{figure}

Figure~\ref{fig:nodes_noncol_diag} and Table~\ref{tab:nodes_noncol_diag} present the corresponding diagonalization times and scaling behavior.
\begin{table}[bt]
  \centering
  \caption{Variation of diagonalization time (s) and scaling with node count for noncollinear DFT. (a) Diagonalization time. (b) Scaling. The system is the same as for Fig.~\ref{fig:nodes_noncol}.}
  \label{tab:nodes_noncol_diag}
  \begin{minipage}[t]{0.48\textwidth}
    \centering
    \begin{tabular}[t]{c c c c}
    \hline
    \rule{0pt}{5ex}
    \begingroup
  \setbox0=\hbox{\shortstack{Number\\ of nodes}}%
  \setbox1=\hbox{\footnotesize(a)}%
  \makebox[\dimexpr \wd1 + 0.60em + \wd0\relax][l]{%
    \raisebox{\dimexpr (\ht0+\dp0-\ht1-\dp1)/2 + 0.10ex\relax}{\copy1}%
    \hspace{0.60em}%
    \copy0
  }%
\endgroup & CPU & \shortstack{GPU-\\accelerated} & \shortstack{Speed-up\\(times)} \\
    \hline
    2 & 1556.7 & 501.3 & 3.11 \\
    4 &  913.3 & 329.8 & 2.77 \\
    6 &  716.6 & 322.3 & 2.22 \\
    8 &  573.8 & 236.2 & 2.43 \\
    \hline
    \end{tabular}
  \end{minipage}
  \hfill
  \begin{minipage}[t]{0.48\textwidth}
    \centering
    \begin{tabular}[t]{c c c}
    \hline
    \noalign{\vskip 0.6ex}
    \begingroup
  \setbox0=\hbox{\shortstack{Number\\ of nodes}}%
  \setbox1=\hbox{\footnotesize(b)}%
  \makebox[\dimexpr \wd1 + 0.60em + \wd0\relax][l]{%
    \raisebox{\dimexpr (\ht0+\dp0-\ht1-\dp1)/2 + 0.10ex\relax}{\copy1}%
    \hspace{0.60em}%
    \copy0
  }%
\endgroup & CPU scaling & GPU scaling \\
    \hline
    2 & 1.00 & 1.00 \\
    4 & 1.70 & 1.52 \\
    6 & 2.17 & 1.56 \\
    8 & 2.71 & 2.12 \\
    \hline
    \end{tabular}
  \end{minipage}
\end{table}
Again, the trends mirror the collinear results, with the GPU speedup decreasing from 3.11 times (2 nodes) to 2.43 times (8 nodes). The performance degradation observed at 6 nodes (6 GPUs) is notable. Since 8 $\mathbf{k}$-points must be processed, 2 GPUs must handle two $\mathbf{k}$-points each, while 4 GPUs handle one each. This load imbalance allows processes with lighter workloads to complete their processing earlier, causing them to enter an idle state while awaiting the completion of processing by the more heavily-loaded processes. Consequently, the overall performance degrades to a level comparable to that of the 4-node case. The limited speedup (1.52 times) when moving from 2 to 4 GPUs is discussed in Sect. 6. The CPU-only configuration shows better scaling (2.71 times from 2 to 8 nodes) than the GPU implementation (2.12 times).

All subsequent results represent the average value for each GPU.

We assess GPU load balance using the coefficient of variation. Table~\ref{tab:load_balance_noncol}(a) shows the results versus number of atoms (8 nodes fixed; 640 atoms excluded due to memory issues). The variation remains below 1\%, indicating excellent balance. Table~\ref{tab:load_balance_noncol}(b) shows the results versus node count (384 atoms fixed; 2 nodes excluded due to memory issues). Except for the 6-node case, the balance is excellent (below 1\%). The high coefficient of variation (approx. 80\%) for 6 nodes confirms the severe workload imbalance discussed previously.
\begin{table}[bt]
  \centering
  \caption{GPU load balance (coefficient of variation) for noncollinear DFT. (a) Variation with number of atoms (8 nodes). (b) Variation with node count (384 atoms). The measurements in panels (a) and (b) are performed on a supercell of diamond-structure silicon.}
  \label{tab:load_balance_noncol}
  \begin{minipage}[t]{0.48\textwidth}
    \centering
    \begin{tabular}[t]{c c c}
    \hline
    \noalign{\vskip 0.6ex}
    \rule{0pt}{9ex}
    \begingroup
  \setbox0=\hbox{\shortstack{Number\\ of atoms}}%
  \setbox1=\hbox{\footnotesize(a)}%
  \makebox[\dimexpr \wd1 + 0.60em + \wd0\relax][l]{%
    \raisebox{\dimexpr (\ht0+\dp0-\ht1-\dp1)/2 + 3.00ex\relax}{\copy1}%
    \hspace{0.60em}%
    \copy0
  }%
\endgroup &
    \shortstack{Coefficient of\\variation for\\kernel time\\only (\%)} &
    \shortstack{Coefficient of\\variation including\\data transfer\\time (\%)} \\
    \hline
    216 & 0.33 & 0.34 \\
    384 & 0.26 & 0.26 \\
    512 & 0.37 & 0.36 \\
    \hline
    \end{tabular}
  \end{minipage}
  \hfill
  \begin{minipage}[t]{0.48\textwidth}
    \centering
    \begin{tabular}[t]{c c c}
    \hline
    \noalign{\vskip 0.6ex}
    \rule{0pt}{9ex}
    \begingroup
  \setbox0=\hbox{\shortstack{Number\\ of nodes}}%
  \setbox1=\hbox{\footnotesize(b)}%
  \makebox[\dimexpr \wd1 + 0.60em + \wd0\relax][l]{%
    \raisebox{\dimexpr (\ht0+\dp0-\ht1-\dp1)/2 + 3.00ex\relax}{\copy1}%
    \hspace{0.60em}%
    \copy0
  }%
\endgroup &
    \shortstack{Coefficient of\\variation for\\kernel time\\only (\%)} &
    \shortstack{Coefficient of\\variation including\\data transfer\\time (\%)} \\
    \hline
    4 & 0.31 & 0.30 \\
    6 & 80.8 & 79.7 \\
    8 & 0.26 & 0.26 \\
    \hline
    \end{tabular}
  \end{minipage}
\end{table}

Table~\ref{tab:gpu_utility_noncol}(a) shows the GPU utilization (defined as in Sect. 5.1) versus number of atoms (8 nodes fixed). Utilization increases with the number of atoms. Table~\ref{tab:gpu_utility_noncol}(b) shows the utilization versus node count (384 atoms fixed). Utilization increases as the node count decreases. These trends mirror the collinear results and are discussed in Sect. 6.
\begin{table}[bt]
  \centering
  \caption{Change in GPU utilization for noncollinear DFT. (a) Variation with number of atoms (8 nodes). (b) Variation with node count (384 atoms). The measurements in panels (a) and (b) are performed on a supercell of diamond-structure silicon.}
  \label{tab:gpu_utility_noncol}
  \begin{minipage}[t]{0.495\textwidth}
    \centering
    \begin{tabular}[t]{c c c}
    \hline
    \noalign{\vskip 0.4ex}
    \rule{0pt}{7ex}
    \begingroup
  \setbox0=\hbox{\shortstack{Number of\\atoms}}%
  \setbox1=\hbox{\footnotesize(a)}%
  \makebox[\dimexpr \wd1 + 0.60em + \wd0\relax][l]{%
    \raisebox{\dimexpr (\ht0+\dp0-\ht1-\dp1)/2 + 0.10ex\relax}{\copy1}%
    \hspace{0.60em}%
    \copy0
  }%
\endgroup & \shortstack{GPU\\utilization (\%)} & \shortstack{GPU utilization\\including\\data transfer\\time (\%)} \\
    \hline
    216 & 5.93 & 6.30 \\
    384 & 14.8 & 15.3 \\
    512 & 22.0 & 22.7 \\
    \hline
    \end{tabular}
  \end{minipage}
  \hfill
  \begin{minipage}[t]{0.495\textwidth}
    \centering
    \begin{tabular}[t]{c c c}
    \hline
    \noalign{\vskip 0.4ex}
    \rule{0pt}{7ex}
    \begingroup
  \setbox0=\hbox{\shortstack{Number of\\nodes}}%
  \setbox1=\hbox{\footnotesize(b)}%
  \makebox[\dimexpr \wd1 + 0.60em + \wd0\relax][l]{%
    \raisebox{\dimexpr (\ht0+\dp0-\ht1-\dp1)/2 + 0.10ex\relax}{\copy1}%
    \hspace{0.60em}%
    \copy0
  }%
\endgroup & \shortstack{GPU\\utilization (\%)} & \shortstack{GPU utilization\\including\\data transfer\\time (\%)} \\
    \hline
    4 & 28.8 & 29.3 \\
    6 & 19.7 & 20.2 \\
    8 & 14.8 & 15.3 \\
    \hline
    \noalign{\vskip 0.6ex}
    \end{tabular}
  \end{minipage}
\end{table}

\begin{figure}[bt]
  \centering
  \begin{minipage}[t]{0.48\textwidth}
    \centering
    \includegraphics[width=\linewidth]{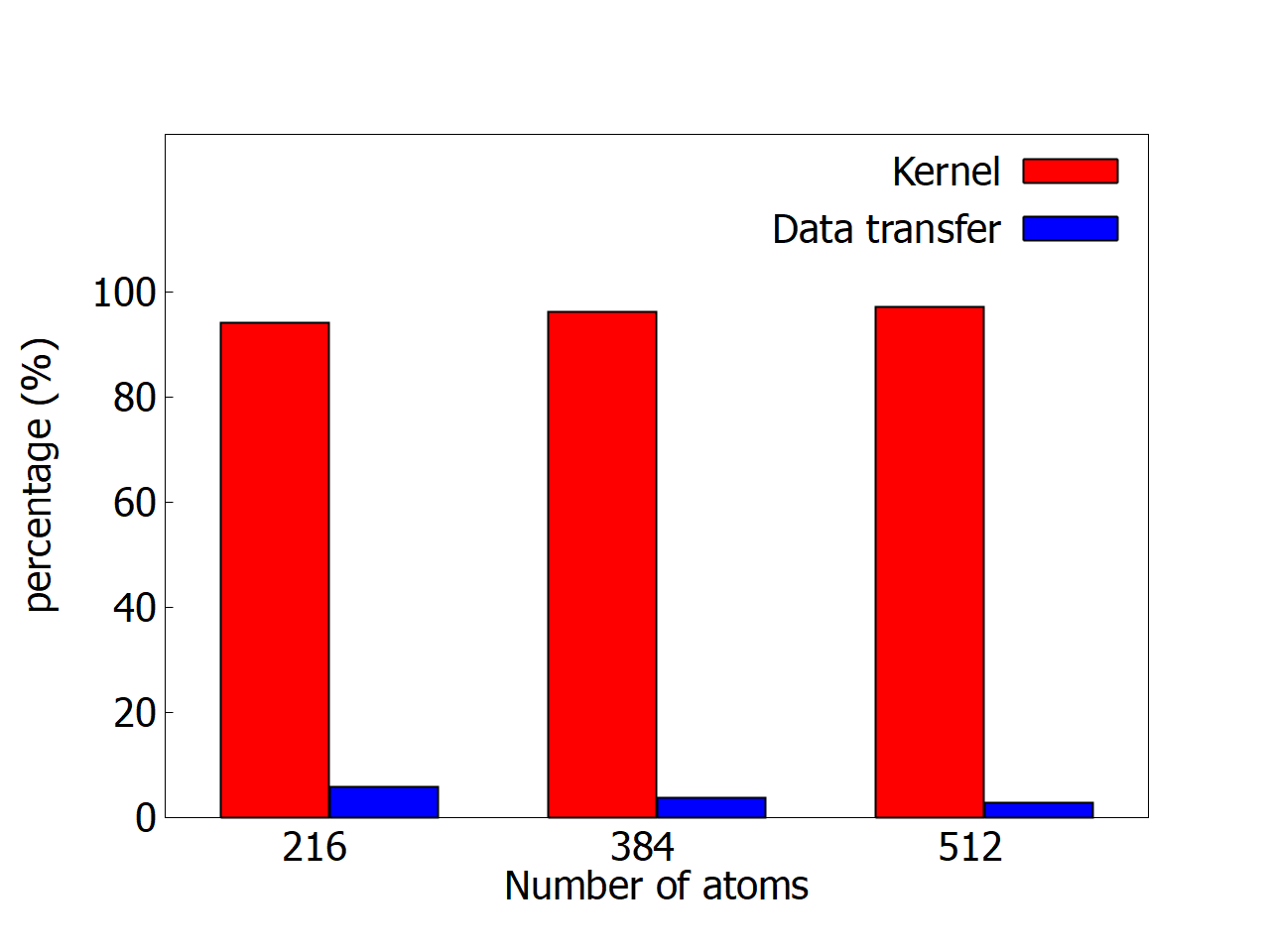}
    \footnotesize{(a)}
  \end{minipage}
  \hfill
  \begin{minipage}[t]{0.48\textwidth}
    \centering
    \includegraphics[width=\linewidth]{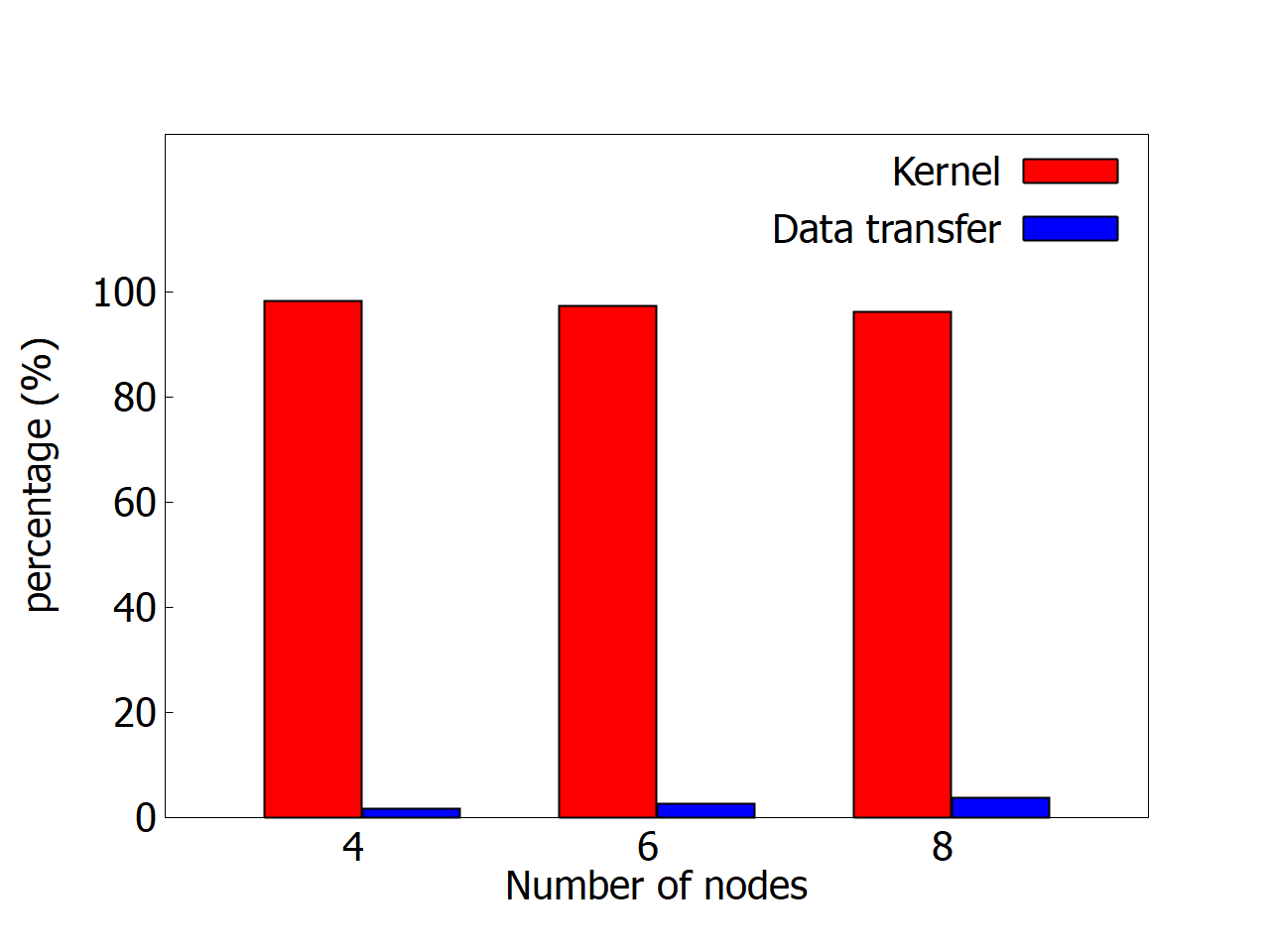}
    \footnotesize{(b)}
  \end{minipage}
  \caption{Fraction of host-device data transfer time in noncollinear DFT. (a) Variation with number of atoms (8 nodes). (b) Variation with node count (384
atoms). The measurements in panels (a) and (b) are performed on a supercell of diamond-structure silicon.}
  \label{fig:percentage_noncol}
\end{figure}

Figure~\ref{fig:percentage_noncol} and Table~\ref{tab:percentage_noncol} analyze the fraction of GPU processing time spent on kernels versus data transfers. Panel (a) shows the variation with number of atoms (8 nodes fixed; 640 atoms excluded). Kernel execution accounts for over 94\% of the time. The kernel fraction increases with number of atoms. Panel (b) shows the variation with node count (384 atoms fixed). Kernel execution again exceeds 95\%. The kernel fraction decreases as the node count increases. These behaviors are similar to the collinear case and are discussed in Sect. 6.

\begin{table}[bt]
  \centering
  \caption{Fraction of host-device data transfer time in noncollinear DFT. (a) Variation with number of atoms. (b) Variation with node count. The measurements in panels (a) and (b) are performed on the same system as in Fig.~\ref{fig:percentage_noncol}.}
  \label{tab:percentage_noncol}
  \begin{minipage}[t]{0.48\textwidth}
    \centering
    \begin{tabular}[t]{c c c}
    \hline
    \rule{0pt}{7ex}
    \begingroup
  \setbox0=\hbox{\shortstack{Number\\ of atoms}}%
  \setbox1=\hbox{\footnotesize(a)}%
  \makebox[\dimexpr \wd1 + 0.60em + \wd0\relax][l]{%
    \raisebox{\dimexpr (\ht0+\dp0-\ht1-\dp1)/2 + 1.20ex\relax}{\copy1}%
    \hspace{0.60em}%
    \copy0
  }%
\endgroup &
    \shortstack{Fraction of\\kernel time (\%)} &
    \shortstack{Fraction of\\data transfer\\time (\%)} \\
    \hline
    216 & 94.18 &  5.82 \\
    384 & 96.33 &  3.67 \\
    512 & 97.13 &  2.87 \\
    \hline
    \end{tabular}
  \end{minipage}
  \hfill
  \begin{minipage}[t]{0.48\textwidth}
    \centering
    \begin{tabular}[t]{c c c}
    \hline
    \rule{0pt}{7ex}
    \begingroup
  \setbox0=\hbox{\shortstack{Number\\ of nodes}}%
  \setbox1=\hbox{\footnotesize(b)}%
  \makebox[\dimexpr \wd1 + 0.60em + \wd0\relax][l]{%
    \raisebox{\dimexpr (\ht0+\dp0-\ht1-\dp1)/2 + 1.20ex\relax}{\copy1}%
    \hspace{0.60em}%
    \copy0
  }%
\endgroup &
    \shortstack{Fraction of\\kernel time (\%)} &
    \shortstack{Fraction of\\data transfer\\time (\%)} \\
    \hline
    4 & 98.41 & 1.59 \\
    6 & 97.33 & 2.67 \\
    8 & 96.33 & 3.67 \\
    \hline
    \end{tabular}
  \end{minipage}
\end{table}

Table~\ref{tab:percentage_noncol_data}(a) details the transfer volume, transfer time, and GPU processing time versus number of atoms (8 nodes fixed).
Table~\ref{tab:percentage_noncol_data}(b) shows these metrics versus the number of nodes (384 atoms fixed; 6 nodes excluded due to imbalance). When reducing from eight nodes (8 GPUs, one $\mathbf{k}$-point per GPU) to four nodes (4 GPUs, two $\mathbf{k}$-points per GPU), the transfer volume per GPU doubles, and the transfer time nearly doubles. However, the GPU processing time increases by nearly a factor of four.
This behavior is similar to that in the collinear calculation, which is discussed in Sect. 6.

\begin{table}[bt]
  \centering
  \caption{Host-device data transfer volume, transfer time, and GPU processing time in noncollinear DFT. (a) Variation with number of atoms. (b) Variation with node count. The measurements in panels (a) and (b) are performed on the same system as in Fig.~\ref{fig:percentage_noncol}.}
  \label{tab:percentage_noncol_data}
  \setlength{\tabcolsep}{4pt}
  \begin{minipage}[t]{0.495\textwidth}
    \centering
    \begin{tabular}[t]{c c c c}
    \hline
    \noalign{\vskip 0.4ex}
    \rule{0pt}{5ex}
    \begingroup
  \setbox0=\hbox{\shortstack{Number\\ of atoms}}%
  \setbox1=\hbox{\footnotesize(a)}%
  \makebox[\dimexpr \wd1 + 0.60em + \wd0\relax][l]{%
    \raisebox{\dimexpr (\ht0+\dp0-\ht1-\dp1)/2 + 1.20ex\relax}{\copy1}%
    \hspace{0.60em}%
    \copy0
  }%
\endgroup &
    \shortstack{Data\\transfer\\volume (MB)} &
    \shortstack{Data\\transfer\\time (sec)} &
    \shortstack{GPU\\processing\\time (sec)} \\
    \hline
    216 &  33433 & 0.950 & 16.3 \\
    384 &  105663 & 2.55 & 69.4 \\
    512 &  187844 & 4.35 & 151.4 \\
    \hline
    \end{tabular}
  \end{minipage}
  \hfill
  \begin{minipage}[t]{0.495\textwidth}
    \centering
    \begin{tabular}[t]{c c c c}
    \hline
    \noalign{\vskip 0.4ex}
    \rule{0pt}{5ex}
    \begingroup
  \setbox0=\hbox{\shortstack{Number\\ of nodes}}%
  \setbox1=\hbox{\footnotesize(b)}%
  \makebox[\dimexpr \wd1 + 0.60em + \wd0\relax][l]{%
    \raisebox{\dimexpr (\ht0+\dp0-\ht1-\dp1)/2 + 1.20ex\relax}{\copy1}%
    \hspace{0.60em}%
    \copy0
  }%
\endgroup &
    \shortstack{Data\\transfer\\volume (MB)} &
    \shortstack{Data\\transfer\\time (sec)} &
    \shortstack{GPU\\processing\\time (sec)} \\
    \hline
    4 & 211326 & 4.65 & 290.8 \\
    8 & 105663 & 2.55 & 69.4 \\
    \hline
    \end{tabular}
  \end{minipage}
\end{table}

\section{Discussions}
The GPU acceleration achieved for collinear DFT (3.27 times speedup in diagonalization for a 1,200-atom system) falls significantly short of the theoretical peak performance ratio between the H100 (51 TFLOPS with Tensor Cores) and the Xeon 8468 (3.2 TFLOPS)—a factor of approximately 16. Several factors contribute to this discrepancy.

First, the serial execution of Part 1 (Fig.~\ref{fig:flowchart_col}) in the GPU implementation introduces a bottleneck absent in the parallelized CPU version. GPU acceleration of Part 1 is non-trivial because efficient parallelization of the required numerical integrations on the GPU is difficult without incurring significant overhead from atomic operations. This remains future work.

Second, although Part 4 is MPI-parallelized, this requires costly redistribution of the matrices (in Part 2-5) from the full representation to the block-cyclic layout. Higher speedups would necessitate GPU acceleration of Part 4 (and Part 3), eliminating the need for redistribution. This is also future work.

Third, the nature of the algorithms limits the attainable speedup. While matrix--matrix multiplication is \textit{compute-bound} and its performance is largely determined by the TFLOPS ratio, eigenvalue computations are often limited by the memory-bandwidth ratio. Matrix multiplication can fully utilize Tensor Cores, approaching the H100's 51 TFLOPS peak. However, eigensolvers utilize matrix multiplications only partially~\cite{gpudc,gpudc2,gpudc3} and cannot fully exploit Tensor Cores, tending toward the non-Tensor-Core peak of 26 TFLOPS.
The memory-bandwidth ratio on Pegasus (GPU: 2~TB/s; CPU: 282~GB/s) is only 7.1 times. Since roughly half of the eigenvalue decomposition algorithm is \textit{bandwidth-bound}, its performance ceiling is expected to be closer to this 7.1 times ratio rather than the 16 times TFLOPS ratio.

For instance, in Part 2-1 (diagonalizing $\mathbf{S}$), where the full spectrum is required, cuSOLVER uses a one-stage Householder reduction, followed by divide-and-conquer~\cite{cusolverhouseholder, cusolvermanual}. The initial Householder reduction (roughly half the workload~\cite{householder}) primarily involves matrix-vector multiplications, which have low arithmetic intensity~\cite{gpuvectormul1, gpuvectormul2} and are \textit{bandwidth-bound}. The subsequent divide-and-conquer stage is dominated by matrix multiplications~\cite{gpudc,gpudc2,gpudc3}, is \textit{compute-bound}, and can exploit Tensor Cores. Consequently, the measured performance ratio (8.15 times) exceeds the memory-bandwidth ratio (7.1 times) but remains below the TFLOPS ratio.

In contrast, for Part 2-4 (diagonalizing $\mathbf{H}'$), typically only one quarter to one half of the spectrum is required. Although cuSOLVER uses the same algorithm for cases where the full spectrum is required and where only a partial spectrum is needed~\cite{cusolvermanual}, the matrix multiplication portion of the divide-and-conquer phase is reduced in the latter case compared to the former. This limits Tensor Core utilization, making the process \textit{bandwidth-bound} and causing the measured performance ratio (4.77 times) to fall below the memory-bandwidth ratio.
The reduced GPU speedup observed for smaller numbers of atoms (smaller matrices) is also significant. With small matrices, the GPU cannot fully exploit its parallelism, leading to idle Streaming Multiprocessors (SMs)~\cite{gpuwake1, gpuwake2}. Furthermore, for smaller systems, the CPU may perform efficiently if the matrix fits into cache or if blocking allows frequent data reuse, hiding main-memory latency~\cite{cpuwake}.

The poor scaling observed when increasing resources [e.g., only 1.44 times improvement when moving from two to four GPUs in the collinear case, Table~\ref{tab:nodes_diag}(b)] is another critical issue. While the serialization of Part 1 contributes to this, the primary factor is the decrease in GPU utilization as the number of nodes increases, as shown in Tables~\ref{tab:gpu_utility}(b) and~\ref{tab:gpu_utility_noncol}(b).

This decrease in utilization occurs because the efficiency gained from sharing a GPU among multiple MPI processes (multiple $\mathbf{k}$-points) is lost when the resources increase. When multiple processes share a GPU (e.g., the two-node configuration), the total GPU processing time increases significantly [e.g., nearly fourfold in Table~\ref{tab:percentage_data}(b)], suggesting concurrent execution of workloads from the MPI processes. However, the active time (time from first GPU event start to last event end) grows less than the total processing time. This is because gaps between kernel executions from one process can be filled by the other process. This overlap effectively hides latency and increases utilization. When the node count increases and each GPU handles fewer $\mathbf{k}$-points (e.g., the four-node configuration), this opportunity for overlap is reduced, leading to lower utilization and poorer scaling. This indicates insufficient work granularity when only a single $\mathbf{k}$-point is processed per GPU.

We also consider the impact of host-device data transfer. As system size increases, the kernel execution time [scaling as $O(N^3)$ for diagonalization] grows much faster than the data transfer time. This is why the fraction of GPU runtime spent on transfers decreases with increasing system size [Tables~\ref{tab:percentage}(a) and~\ref{tab:percentage_noncol}(a)], contributing to better performance for larger systems.
However, since the transfer ratio is 14\% or less, it is not a primary factor for performance improvement in large-scale systems.

Regarding scaling with node count, both the computational workload and the data transfer volume per GPU are proportional to the number of $\mathbf{k}$-points handled by that GPU. When increasing the node count from two to four [Table~\ref{tab:percentage_data}(b)], the transfer time per GPU decreases (from 4.42 to 2.31 s). However, the kernel execution time decreases even more rapidly. Consequently, the fraction of time spent on data transfer increases [from 3.65 to 7.90\%, Table~\ref{tab:percentage}(b)]. This relative increase in transfer overhead contributes to the degradation of scaling performance. Nevertheless, since the transfer fraction remains below 8\%, it is not the dominant factor limiting scalability. We note that data transfers from multiple processes sharing a GPU appear to be effectively serialized, either through time-slicing or asynchronous execution.

For small systems, the overall DFT speedup is also limited because non-diagonalization procedures (e.g., \texttt{Set Hamiltonian}) constitute a larger fraction of the runtime. Accelerating these procedures on the GPU is necessary for further performance gains in small systems.

These considerations apply similarly to noncollinear DFT. However, the speedup for diagonalizing $\mathbf{S}$ (Part 3-1) is only 5.44 times, falling below the memory-bandwidth ratio. This lower performance compared to the collinear case (8.15 times) is because the noncollinear benchmarks involve fewer atoms (due to host memory constraints), resulting in smaller matrices where the effects of insufficient granularity and transfer overhead are more pronounced.
A limitation of using cuBLAS and cuSOLVER in this manner is that the full-size matrix must fit into the device memory, and the matrix dimension must be 32,768 or less (an API limitation). In collinear DFT, an $n \times n$ complex matrix (where $n$ is the number of basis functions) occupies $16n^2$ bytes (FP64). The CUDA 12.0 eigenvalue solver requires approximately three times this memory (matrix plus workspace). On an 80 GB H100 GPU, the memory constraint allows for approximately 42,000 basis functions. However, in practice, the API constraint limits the system to 32,768 basis functions.

This limit (32,768 basis functions) corresponds to approximately 2,500 silicon atoms using the basis set in this study. For noncollinear DFT, the matrix dimension is $2n$, occupying $64n^2$ bytes. The memory constraint limits this to approximately 21,000 basis functions. The API constraint further limits $2n$ to 32,768, meaning $n$ is limited to 16,384 basis functions (approximately 1,300 silicon atoms). These limits cover many practical calculations.
For larger systems exceeding these limits, the matrices cannot be stored on a single GPU. 
To address such cases, one approach is to utilize distributed-memory parallelization across multiple GPUs with MPI-enabled GPU libraries such as cuBLASMp~\cite{cublasmp} and ELPA GPU~\cite{elpagpu1,elpagpu2}.
An advantage of the GPU-accelerated implementation is its memory efficiency regarding MPI processes. In CPU-only OpenMX, host-memory constraints sometimes prevent assigning one MPI process per core, limiting performance. With GPU acceleration, fewer MPI processes can be used without degrading GPU performance, enabling high execution speed even when host memory is limited.

\section{Conclusions}
We implemented GPU acceleration for both collinear and noncollinear DFT in the NAO-based code OpenMX. For a 512-atom collinear system on two nodes (two GPUs), we achieved a 2.02 times speedup relative to a CPU-only configuration (96 cores). For a 384-atom noncollinear system under the same configuration, we achieved a 2.60 times speedup. These gains represent only about one-sixth of the theoretical GPU-to-CPU TFLOPS ratio ($\sim$16 times).

This discrepancy was primarily because only dense matrix multiplications and eigenvalue solvers were offloaded to the GPU. While matrix multiplications achieved speedups approaching the theoretical TFLOPS ratio, the eigenvalue computations were limited by the memory-bandwidth ratio (7.1 times).

Furthermore, forced serialization of preceding steps (Hamiltonian matrix and overlap matrix construction) and the overhead of matrix redistribution for subsequent MPI-parallelized steps created bottlenecks.
Further acceleration requires offloading additional parts of the code to the GPU.
Moreover, the utilization of distributed GPU algorithms becomes crucial for addressing ultra-large-scale systems.

\section*{Acknowledgments}
We sincerely thank Dr. Shinnosuke Furuya of NVIDIA for his valuable suggestions regarding GPU acceleration. We also acknowledge the advice and insights received during the “GPU Mini Camp” held at the Information Technology Center, University of Tokyo. 
The authors acknowledge the assistance of AI-based language models, ChatGPT (OpenAI) and Gemini (Google), for improving the language and clarity of this manuscript. The authors are solely responsible for the content and any remaining errors.
This research (in part) used computational resources of Pegasus provided by Multidisciplinary Cooperative Research Program in Center for Computational Sciences, University of Tsukuba.
The GPU-accelerated OpenMX test calculations were performed using the supercomputer HPE SGI8600 at the Japan Atomic Energy Agency and the computational resources of the Research Center for Computational Science, Okazaki, Japan (Project: 23-IMS-C987).
This work was partially supported by JSPS KAKENHI Grant Number 25K07201 and Joint Usage Program of the Institute for Solid State Physics, The University of Tokyo (Project No. 202305-VSBXS-0007).

\end{document}